\documentclass[12pt]{article}
\usepackage{amssymb}
\usepackage{graphicx}
\usepackage{graphics}
\usepackage{epsfig}
\usepackage{setspace,caption}
\makeatletter
\newcommand{\row}[1]{\mathord{\buildrel{\lower3pt\hbox{$\scriptscriptstyle\rightarrow$}}\over #1}}

\newcommand{\dyadic}[1]{\mathord{\dyadic@rrow{#1}}}
\newcommand{\dyadic@rrow}[1]{
\begin{picture}(12,12)(-1,0)
\put(-2,10){\makebox(0,0)[t]{$\scriptscriptstyle\downarrow$}}
\put(-2,11){\makebox(0,0)[l]{$\scriptscriptstyle\longrightarrow$}}
\put(5,0){\makebox(0,0)[b]{$#1$}}
\end{picture}
}
\newcommand{\bra}[1]{\bigl\langle #1 \bigr|}
\newcommand{\ket}[1]{\bigl| #1 \bigr\rangle}

\begin{document}

\begin{center}
\textbf{ Fisher information of  a single qubit interacts with a
spin-qubit in the presence of a magnetic field }
 \vspace{0.5 cm}\\ {\small
N. Metwally\\\footnote[1]{E-mail: nmetwally@uob.edu.bh} {\small
Department of Mathematics, College of Science, University of
Bahrain, Bahrain.}

{\small Department of Mathematics,  Faculty of Science, Aswan
University, Egypt}}

\end{center}

 {\bf Abstract:} {In this contribution, quantum Fisher information is utilized
  to estimate the  parameters of a central  qubit interacting with a single-
 spin qubit.   The effect of the longitudinal, transverse
  and the rotating strengths of the  magnetic field  on the estimation degree  is
 discussed. It is shown that, in the resonance case, the number of peaks and
consequently the  size of the  estimation regions  increase as the
rotating magnetic field strength increases. The precision
estimation of the central qubit parameters depends on the initial
state settings of the central  and the spin- qubit, either encode
classical or quantum information. It is displayed that, the upper
bounds of the estimation degree are large
 if the two qubits  encode classical information.  In the non-resonance case, the estimation
degree depends on which of  the longitudinal/transverse strength
is larger. The coupling constant   between the central qubit and
the spin- qubit  has a different effect on  the estimation degree
of the weight and the phase parameters, where the possibility of
estimating the weight parameter decreases as the coupling constant
increases, while it increases for the phase parameter.

For large number of spin-particles, namely, we have  a spin-bath
particles, the  upper bounds of  the Fisher information  with
respect to the weight parameter of the central qubit  decreases as
the number of the spin particle increases. As the interaction time
increases, the upper bounds appear at different initial values of
the weight parameter.
  }

\section{Introduction}
Fisher information  plays an  important  role in the context of
quantum metrology \cite{Holevo} and quantum information processing
\cite{ Sun2010, Ming2010, Nan2013, Metwally2016, Metwally017}.
Quantum Fisher information (QIF) quantifies the information that
can be elicited about a parameter. In other words, QFI is used as
an estimation tool  of parameters that contained in  the quantum
system during its evolution \cite{Ma2011}.  Due to its importance
, there are some efforts  that has been done to quantify QFI in
different quantum systems.

Recently, quantifying  quantum Fisher information in open quantum
systems has paid attentions by some authors. For example, Zheng
et. al \cite{Zheng} investigated the dynamics of QFI for a
two-qubit system, where each qubit interacts with its own
Markovian environment. Ozaydin  \cite{Fathi} has quantified the
QFI analytically  for  the W- state  in the presence of different
noisy channels. The effect of the Markovian reservoirs on the
dynamics of the quantum Fisher information of  a two-level system
is discussed by G.-You et.al \cite{You}.
 Quantum Fisher information for a noisy
open quantum system and  initially prepared in a steady state is
quantified by Altinats \cite{Altinats}.

The central-spin system represents one of the most important
models of decoherence\cite{Breuer}. Rao \cite{Rao} discussed the
dynamics of one and two-qubit systems interacting with a spin-
qubit. The coherent and the  non-coherent  properties of a single
and a maximum entangled two-qubit systems interact with a spin-
qubit are discussed by Metwally et. al \cite{Metwally013}. The
dynamics of quantum Fisher information for a spin-boson model is
investigated by   Hao et. al \cite{Hao}. The dynamics of the
quantum Fisher information for a qubit system and   initially
prepared in a coherent spin-squeezing state is investigated  by
Zhong et. al. \cite{Wei}.

 In the present work, we   quantify the  quantum Fisher information  of a
 single  central  qubit interacting with a single spin-qubit in the presence of a magnetic field. The QFI
 is utilized to estimate the weight and the phase  parameters which
 describe  the initial state of the central qubit. The paper is organized as
 follows. In  Sec.$2$, we introduce the suggested model, where an
 analytical solution in terms of the Bloch vector is given.
 A brief description of  quantum Fisher  information is given in
 Sec.$3$. The effect of longitudinal and transverse strengths  of the magnetic  field
 on the
estimation degree of the weight parameter is given in Sec.$3$.  In
Sec.$4$, the behavior of the quantum Fisher information with
respect to the phase parameter is described. In Sec. 5, we
investigate the behavior of the Fisher information with respect to
the weight parameter for large numbers of spin-bath particles.
Finally, we draw our conclusion in Sec.$5$.

 \section{Presentation of the model}

The Hamiltonian operator which describes the interaction of a
central  single 2-level qubit with upper and ground states
$\ket{\pm}$ with a single- spin $\frac{1}{2}$ particle is given by
\cite{Rao},
\begin{equation}
\hat H=\hbar \omega _{0}\hat\sigma _{z}+\hbar \omega _{1}(\hat
S_{+}e^{-i\omega t}+\hat S_{-}e^{i\omega t})+\hat S_{z}g,
\end{equation}%
where the spin operators $\hat S_{\pm,z}$ satisfy $\Bigl[\hat
S_z,\hat S_{\pm}\Bigr]=\pm\hat S_{\pm}$,  $\Bigl[\hat S_{+},\hat
S_{-}\Bigr]=2\hat S_z$. The parameters  $\omega_0$ and $\omega$,
represent the longitudinal and the transverse components of the
magnetic  field, respectively, $\omega_1$ is the rotating
 magnetic field strength and $g$ is the coupling constant between the spin-qubit and the central qubit.
  In this treatment, it is assumed that, the initial
state of the system is described by the product state
$\rho_s(0)=\rho_q(0)\otimes\rho_b(0)$, where
$\rho_q(0)=\ket{\psi_q}\bra{\psi_q}$ and
$\rho_b(0)=\ket{\chi_b}\bra{\chi_b}$ represent the states of the
central qubit and the spin-qubit, respectively. In the
computational basis $\{\ket{0}, \ket{1}\}$, the  states
$\ket{\psi_q}$ and $\ket{\chi_b}$  may be written as,
\begin{eqnarray}
\ket{\psi_q(0)}&=&\cos\theta_1\ket{0}+\sin\theta_1e^{-i\phi_1}\ket{1},
\nonumber\\
\ket{\chi_b(0)}&=&\cos\theta_2\ket{0}+\sin\theta_2e^{-i\phi_2}\ket{1},
\end{eqnarray}
where $(\theta_1,\phi_1)$ and $(\theta_2,\phi_2)$ are the weight
and the phase  angles of the central and the spin- qubits,
respectively. The  Bloch vectors of these initial states are given
by,
\begin{eqnarray}
s_{x_i}(0)&=&\sin\theta_i\cos\phi_i,\quad
s_{y_i}(0)=\sin\theta_i\sin\phi_i,\quad s_{z_i}(0)=\cos\theta_i ,
i=1,2.
\end{eqnarray}
At any $t>0$, the final state of the total system is given by,
\begin{equation}
\ket{\psi_{qb}(t)}=\mu(t)\ket{\psi_{qb}(0)},\quad
~\mu(t)=e^{-i\hat Ht}.
\end{equation}
In the computational basis set, $\{00,01,10$  $11$\}, the  unitary
operator  $\mu(t)$ may be described by the following $4\times 4$
matrix,
\begin{equation}
\mathcal{\mu}(t)= \left(
\begin{array}{cccc}
\mathcal{\mu}_{11}&0&\mathcal{\mu}_{13}&0
\\
0&\mathcal{\mu}_{22}&0&\mathcal{\mu}_{24}\\
 \mathcal{\mu}_{31}&0&\mathcal{\mu}_{33}&0\\
0& \mathcal{\mu}_{42}&0&\mathcal{\mu}_{44}
\end{array}
\right),
\end{equation}
where
\begin{eqnarray}
\mathcal{\mu}_{11}&=&\frac{2i\omega _{1}}{\gamma _{2}}e^{i\omega
t/2}\sin \gamma _{2}t,\quad
\mathcal{\mu}_{13}=e^{-i\omega t/2}\bigl(\cos \gamma _{2}t-2i\frac{\Delta _{-}%
}{\gamma _{2}}\sin \gamma _{1}t\bigr),
\nonumber\\
\mathcal{\mu}_{22}&=&e^{-i\omega t/2}\bigl(\cos \gamma _{1}t-\frac{2i\Delta _{+}%
}{\gamma _{1}}\sin \gamma _{1}t\bigr),\quad
\mathcal{\mu}_{24}=\frac{2i\omega _{1}}{\gamma _{1}}e^{-i\omega
t/2}\sin \gamma _{1}t,
\nonumber\\
\mathcal{\mu}_{31}&=&\frac{2i\omega _{1}}{\gamma _{2}}e^{i\omega
t/2}\sin \gamma _{2}t, \quad \mathcal{\mu}_{33}=e^{i\omega t/2}\bigl(\cos \gamma _{2}t+2i\frac{\Delta _{-}%
}{\gamma _{2}}\sin \gamma _{1}t\bigr)
\nonumber\\
\mathcal{\mu}_{42}&=&\frac{2i\omega _{1}}{\gamma _{1}}e^{i\omega
t/2}\sin \gamma _{1}t,\quad\mathcal{\mu}_{44}=e^{i\omega t/2}\bigl(\cos \gamma _{1}t+\frac{2i\Delta _{+}%
}{\gamma _{1}}\sin \gamma _{1}t\bigr),
\end{eqnarray}
and  $\Delta _{\pm}=\Delta \pm\frac{g}{2},$  $\gamma _{1}=\sqrt{\omega _{1}^2+\Delta _{+}^2},$ $\gamma _{2}=\sqrt{%
\omega _{1}^2+\Delta _{-}^2}$ and $\Delta =\omega -\omega _{0}$ is
the detuning parameter.

By using Eqs.(2-6), the state (4) may be written as,
\begin{equation}
\ket{\psi_{qb}(t)}=\mathcal{B}_1\ket{00}+\mathcal{B}_2\ket{01}+\mathcal{B}_3\ket{10}+\mathcal{B}_4\ket{11},
\end{equation}
where,
\begin{eqnarray}
\mathcal{B}_1&=&\mu_{11}c_1c_2+\mu_{13}s_1c_2e^{-i\phi_1}, ~\quad
\mathcal{B}_2=\mu_{22}c_1s_2e^{-i\phi_2}+\mu_{24}s_1s_2e^{-i\phi_{12}} \nonumber\\
\mathcal{B}_3&=&\mu_{31}c_1c_2+\mu_{33}s_1c_2e^{-i\phi_1}, ~\quad
\mathcal{B}_4=\mu_{42}c_1s_2e^{-i\phi_2}+\mu_{44}s_1s_2e^{-i\phi_{12}},
\end{eqnarray}
and $c_i=\cos\theta_i$, $s_i=\sin\theta_i$, $i=1,2$ and
$\phi_{12}=\phi_1+\phi_2$. Since we are interested in
investigating the dynamics of the Fisher information with respect
to the central qubit, we trace out the state of the spin- qubit,
where $\rho _{q}(t)=tr_{b}\{\rho _{s}(t)\}$
  and $\rho_q(t)=\ket{\psi_{qb}(t)}\bra{\psi_{qb}(t)}$. In the Bloch vector representation, the  state
   $\rho_q(t)$ can be written as,
   \begin{equation}
   \rho_q(t)=\frac{1}{2}(1+s_{x_q}(t)\sigma_x+s_{y_q}\sigma_y(t)+s_{z_q}(t)\sigma_z),
   \end{equation}
where
\begin{eqnarray}
s_{x_q}(t)&=&\mathcal{B}_1\mathcal{B}^*_3+\mathcal{B}_3\mathcal{B}^*_1+\mathcal{B}_2\mathcal{B}^*_4+\mathcal{B}_4\mathcal{B}^*_2,
\nonumber\\
s_{y_q}(t)&=&i(\mathcal{B}_3\mathcal{B}^*_1+\mathcal{B}_2\mathcal{B}^*_4-\mathcal{B}_1\mathcal{B}^*_3-\mathcal{B}_4\mathcal{B}^*_2),
\nonumber\\
s_{z_q}(t)&=&|\mathcal{B}_1|^2+|\mathcal{B}_2|^2-|\mathcal{B}_3|^2-|\mathcal{B}_4|^2,
\end{eqnarray}
where $\mathcal{B}_i, i=1...4$ are given by (8).

\section{Quantum Fisher Information}
In the following subsection, we review the mathematical form of
the  quantum Fisher information for a single qubit in the Bloch
vector representations. Moreover, we quantify numerically
 the QFI corresponding to the weight and the phase
parameters of the central qubit.
\subsection{Mathematical form }
The density operator of a single qubit is given by,
\begin{equation}
\rho=\frac{1}{2}(I+\sum_i^3{s_i\sigma_i}),
\end{equation}
where $I$ is a unit matrix of size $2\times 2$,
$\row{s}=(s_1,s_2,s_3)$ is the Bloch vector and $\sigma_i,
i=x,y,z$ are the Pauli matrices.
 In terms of the Bloch vector,  the QFI with respect
to a  parameter $\eta$ is defined as \cite{Wei013,Xing016},

\begin{equation}
\mathcal{F}_{\eta} = \left\{ \begin{array}{ll}
\frac{1}{1-\bigl|\row{s}(\eta)\bigr|^2}\Bigl[\row{s}(\eta)\cdot\frac{\partial{\row{s}(\eta)}}{\partial\eta}\Bigr]
+\Bigl(\frac{\partial\row{s}(\eta)}{\partial\eta}\Bigr)^2&,  \textrm{ for mixed state},|\row{s}(\eta)|<1,\\
\nonumber\\
\Bigl|\frac{\partial\row{s}(\eta)}{\partial\eta}\Bigr|^2 &, \quad\textrm{for pure state}, ~|\row{s}(\eta)|=1,\\
\end{array} \right.
\end{equation}
where $\eta$ is the parameter to be estimated.
 In the following subsections, we shall estimate the weight  and the phase parameters of the central qubit, where it is assumed that
 the spin qubit is polarized in $z-$ direction, namely, $\rho_b=\frac{1}{2}(1+\sigma_z)$. For this aim, we calculate
the quantum Fisher information, $\mathcal{F}_\eta$ corresponding
to the weight and the phase parameters, i.e.,
$\eta=\theta_1,\phi_1$.

\begin{figure}[b!]
\begin{center}

\includegraphics[width=16pc,height=12pc]{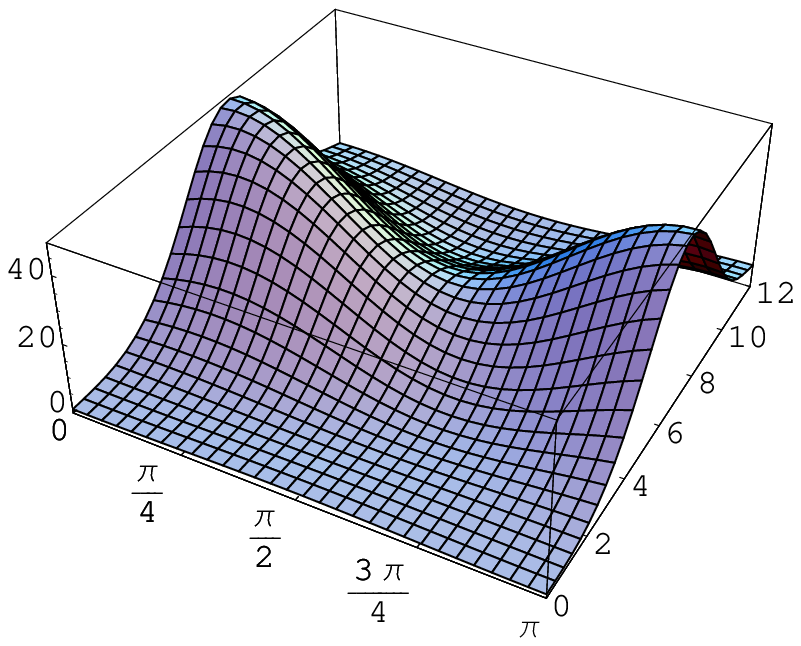}
\put(-140,8){\large $\theta_1$}\put(-215,70){\large
$\mathcal{F}_{\theta_1}$} \put(-20,35){\large $t$}
\put(-180,150){(a)} ~~\quad
\includegraphics[width=16pc,height=12pc]{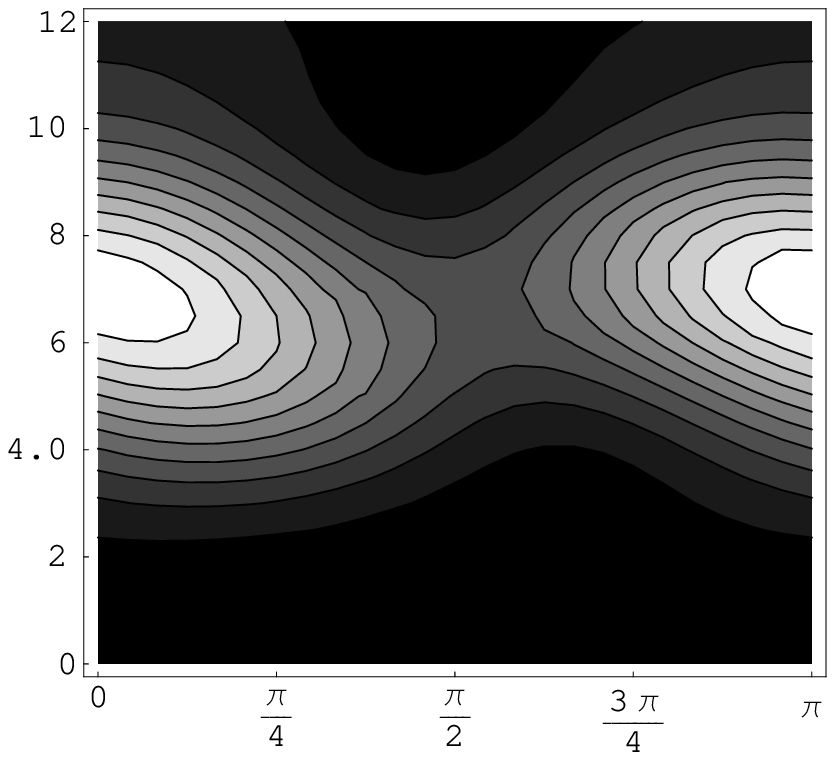}
\put(-100,-8){\large $\theta_1$}\put(-190,70){\large $t$}
\put(-20,150){(b)}~~\quad\\
\includegraphics[width=16pc,height=12pc]{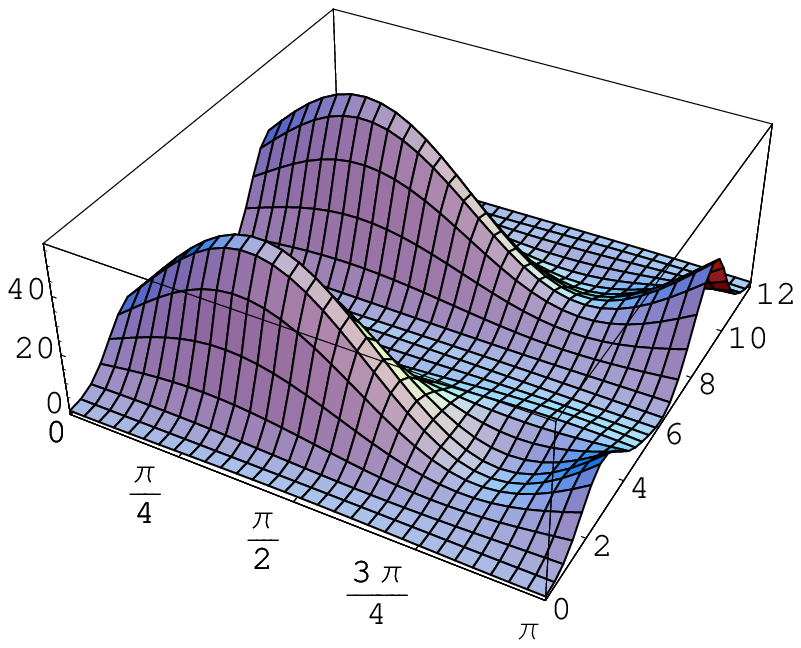}
\put(-140,8){\large $\theta_1$}\put(-215,70){\large
$\mathcal{F}_{\theta_1}$}\put(-20,35){\large $t$}
\put(-180,150){(c)}~~\quad
\includegraphics[width=16pc,height=12pc]{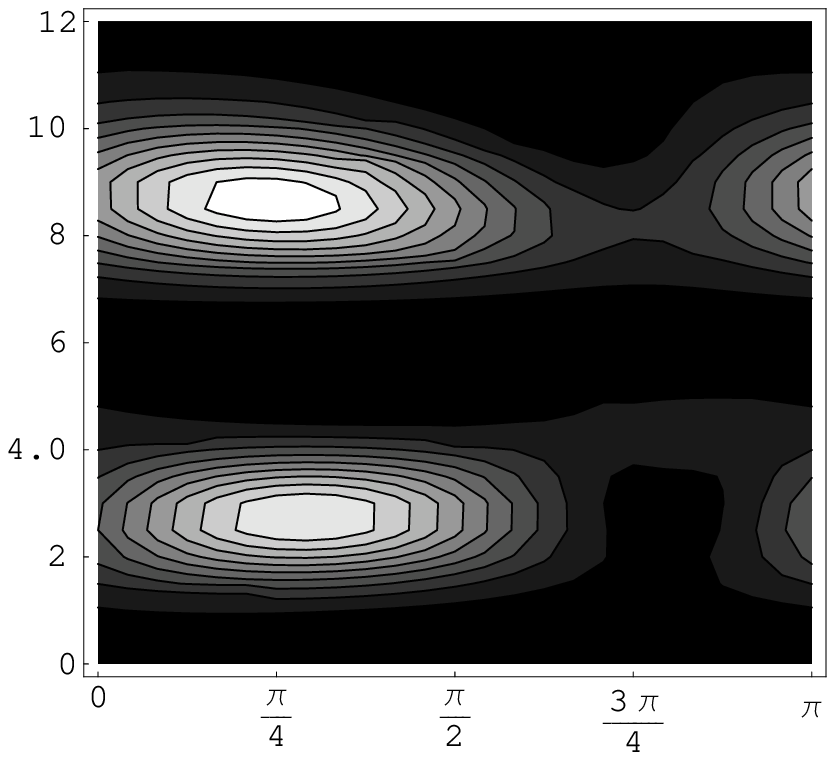}
\put(-100,-8){\large $\theta_1$}\put(-190,70){\large $t$}
\put(-20,150){(d)}\\
\includegraphics[width=16pc,height=12pc]{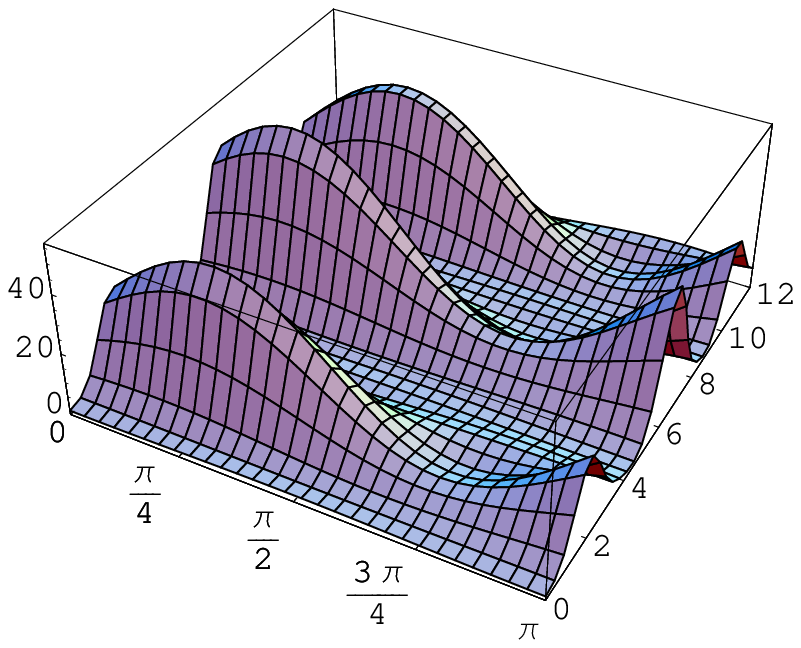}
\put(-140,8){\large $\theta_1$}\put(-215,70){\large
$\mathcal{F}_{\theta_1} $}\put(-20,35){\large
$t$}\put(-180,150){(e)}~~\quad
\includegraphics[width=16pc,height=12pc]{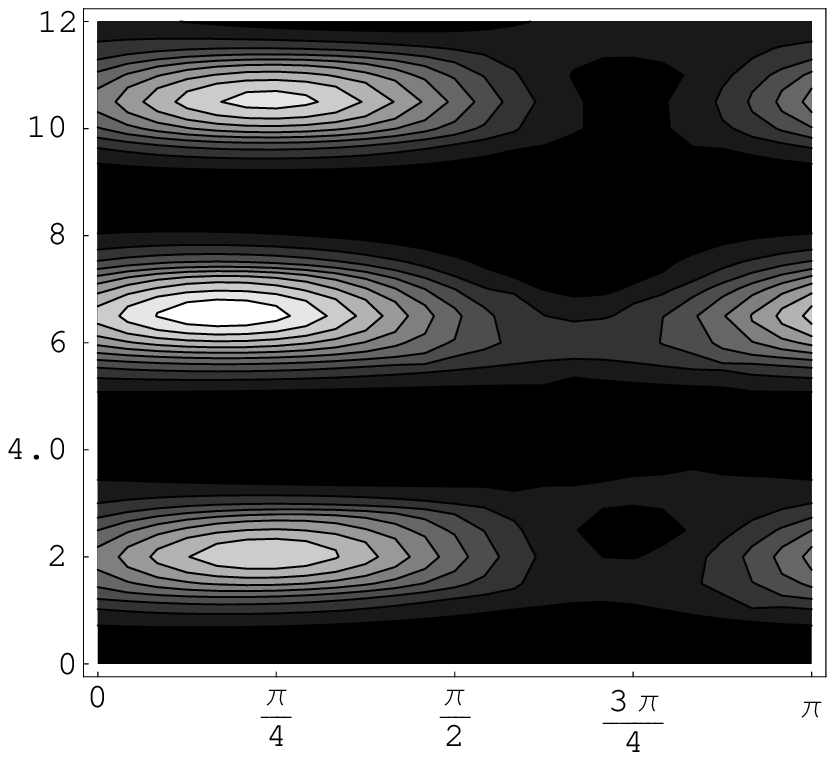}
\put(-100,-8){\large $\theta_1$}\put(-190,70){\large $t$}\put(-20,150){(f)}\\

\end{center}
\caption{ Fisher information  for resonance case, namely,
$\Delta=\omega-\omega_0=0$, where $g=0.5$, $\phi=\pi$ and
$\omega_1=0.1,0.5,7$, for (a,b),(c,d) and (e,f), respectively. }
\end{figure}

\subsection{Numerical results}
We assume  that, the phase angle $\phi_1=\pi$, namely, the initial
state of the central qubit is prepared in the state
$\ket{\psi_q}=\cos\theta_1\ket{0}-\sin\theta_1\ket{1}$, while the
spin-qubit is prepared in the state $\ket{\chi(0)}=-\ket{0}$,
namely $\theta_2=\pi$ and the phase $\phi_2$ is arbitrary.

Fig.(1) describes the dynamics of  the  quantum Fisher information
($\mathcal{F}_{\theta_1})$ with respect to the weight parameter
$(\theta_1)$. The effect of the rotating  magnetic field strength
$(\omega_1)$ on the quantum Fisher information
$\mathcal{F}_{\theta_1}$ is investigated in the resonance case (
$\Delta=0$) and  for a fixed value of the coupling constant
between the central and spin qubits $(g=0.5)$. The general
behavior shows that, the maximum values of Fisher information
depend on the initial weight angle $(\theta_1)$ and the
interaction time.  As it is displayed in Figs.(1a) and (1b), the
Fisher information $\mathcal{F}_{\theta_1}$ increases gradually to
reach its maximum value at $t=0.7$. For further values of  the
interaction time, $\mathcal{F}_{\theta_1}$ decreases gradually.
The effect of the initial weight settings shows that, Fisher
information decreases gradually to its minimum value at
$\theta_1=\pi/2$, namely the initial state of the central qubit is
papered in the state $\ket{\psi_q(0)}=\ket{1}$. However, the
behavior changes for  $\theta_1\in[\pi/2,\pi]$, where the Fisher
information increases to reach its maximum bounds at
$\theta_1=\pi$.

The effect of  larger values of the rotating  magnetic field
strength $\omega_1$  is depicted in Fig.s(1c-1f), where the number
of peaks increases as $\omega_1$ increases and the maximum bounds
appear at larger values of $\theta_1$. On the other hand, the
larger values of $\omega_1$ has no effect on the upper bounds of
$\mathcal{F}_{\theta_1}$. Moreover,  as the strength of the
rotating magnetic field strength  $(\omega_1)$ increases, the
quantum Fisher information increases suddenly to reach its maximum
values. The contour description( Figs.1(b,d,f))  shows that,  the
size of bright regions in which one may estimate the weight
parameter, decreases as the initial weight parameter $\theta_1$
increases. As the rotating field strength $(\omega_1)$ increases,
the bright regions are shifted as $\theta_1$ increases. The number
of peaks increases at  the expense of the  bright  size regions.

\begin{figure}[b!]
\begin{center}

\includegraphics[width=16pc,height=12pc]{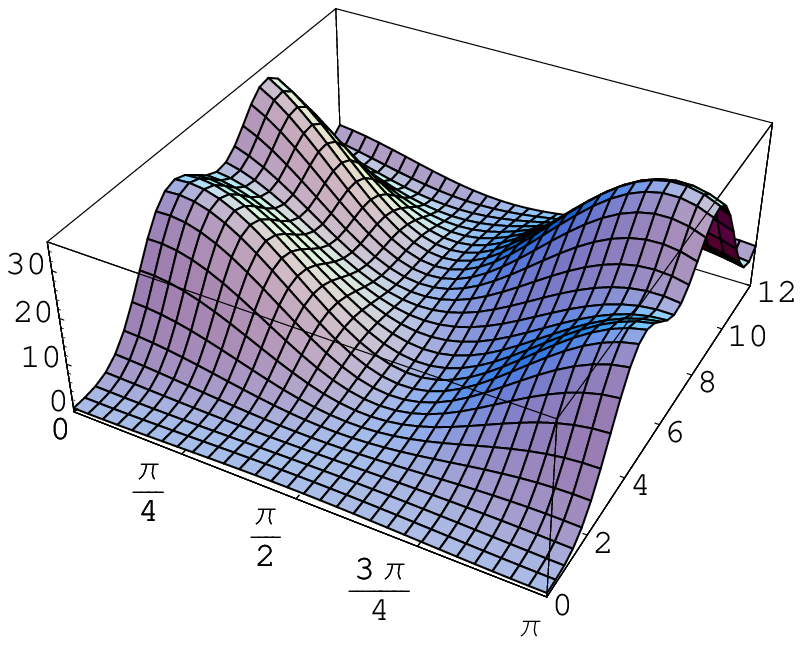}
\put(-215,70){\large $\mathcal{F}_{\theta_1}$}
\put(-145,12){\large$\theta_1$} \put(-20,40){\large
$t$}\put(-180,150){(a)}~~\quad
\includegraphics[width=16pc,height=12pc]{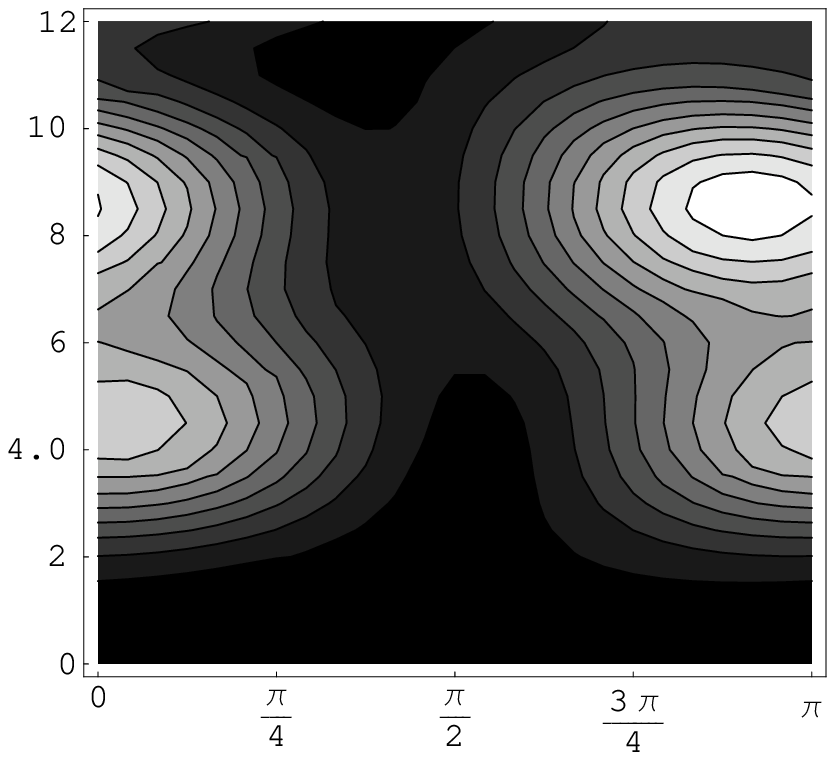}
\put(-100,-8){\large $\theta_1$} \put(-190,70){\large $t$}\put(-20,150){(b)}\\
\includegraphics[width=16pc,height=12pc]{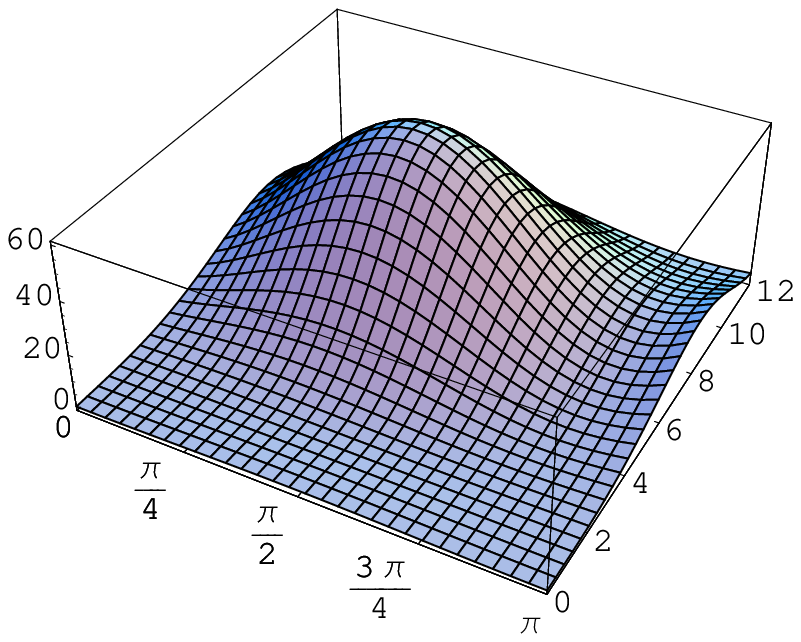}
\put(-215,70){\large $\mathcal{F}_{\theta_1}$}
\put(-145,12){\large$\theta_1$} \put(-20,40){\large
$t$}\put(-180,150){(c)}~~\quad
\includegraphics[width=16pc,height=12pc]{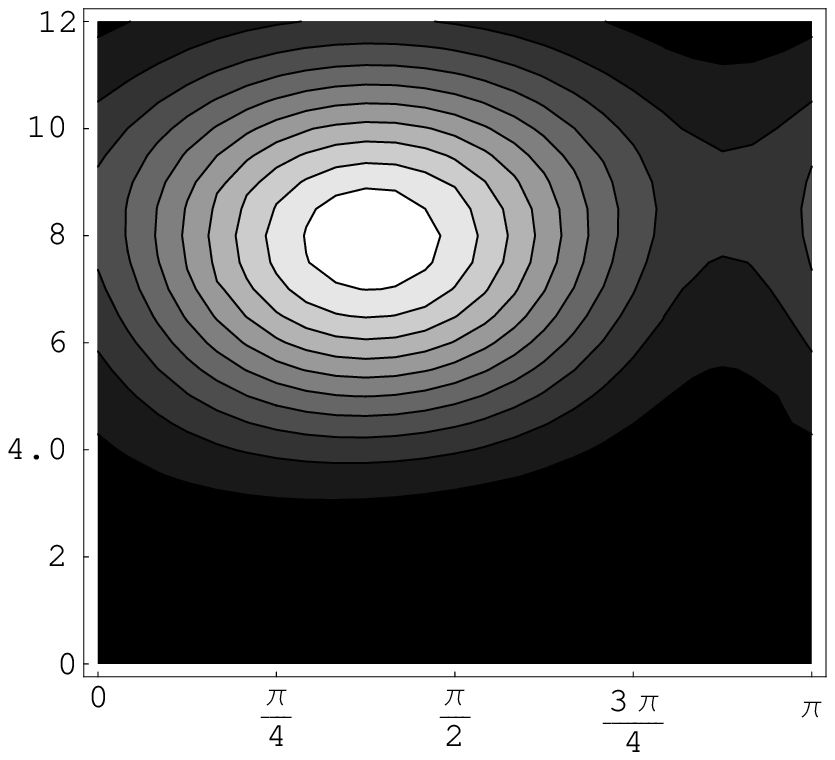}
\put(-100,-8){\large $\theta_1$} \put(-190,70){\large
$t$}\put(-20,150){(d)}
\end{center}
\caption{ For the  non-resonance case where $\Delta=-0.08,0.08$
namely we set $\omega_0=0.01, \omega=0.09$ in (a,b) and
$\omega_0=0.09, \omega=0.01$ in (c,d), where $\omega=0.1. g=0.5,
\phi=\pi$.}
\end{figure}

\begin{figure}[b!]
\begin{center}
\includegraphics[width=16pc,height=12pc]{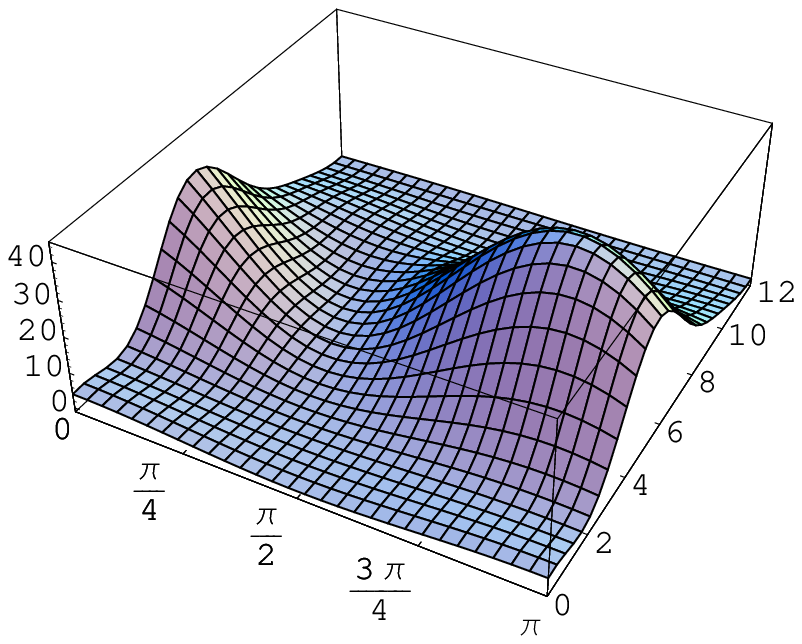}
\put(-215,70){\large $\mathcal{F}_{\theta_1}$}
\put(-145,12){\large$\theta_1$} \put(-20,40){\large
$t$}\put(-180,150){(a)}~~\quad
\includegraphics[width=16pc,height=12pc]{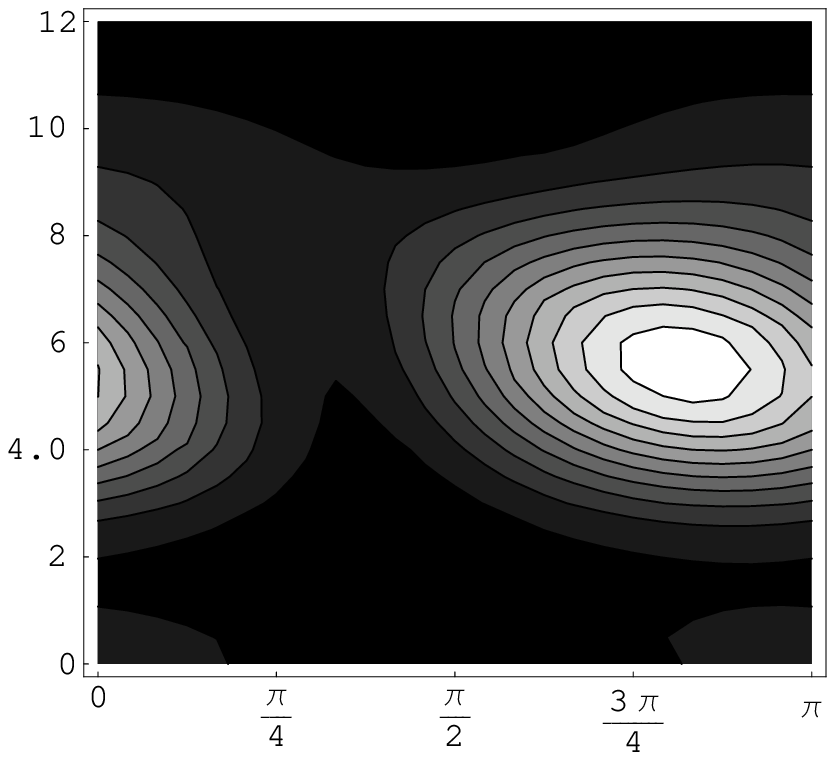}
\put(-100,-8){\large $\theta_1$} \put(-190,70){\large $t$}\put(-20,150){(b)}\\
\includegraphics[width=16pc,height=12pc]{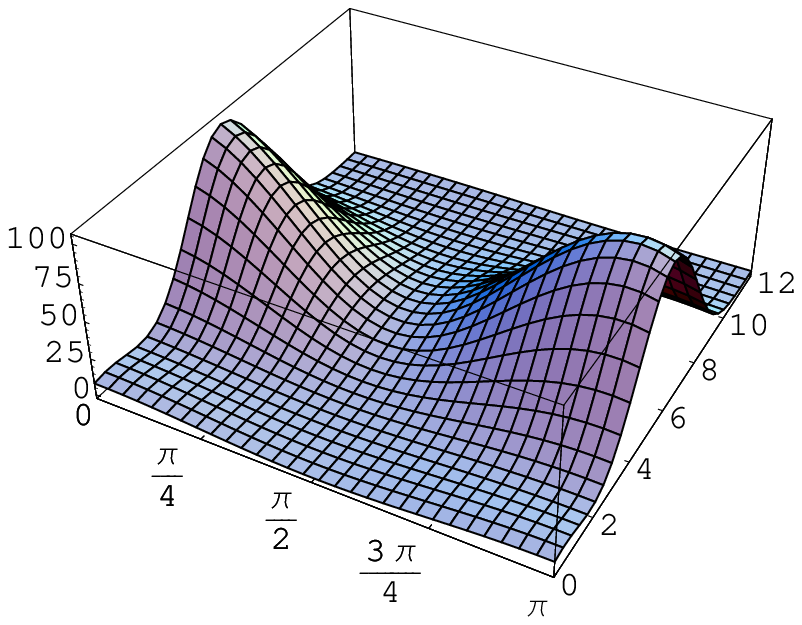}
\put(-215,70){\large $\mathcal{F}_{\theta_1}$}
\put(-145,12){\large$\theta_1$} \put(-20,40){\large
$t$}\put(-180,150){(c)}~\quad
\includegraphics[width=16pc,height=12pc]{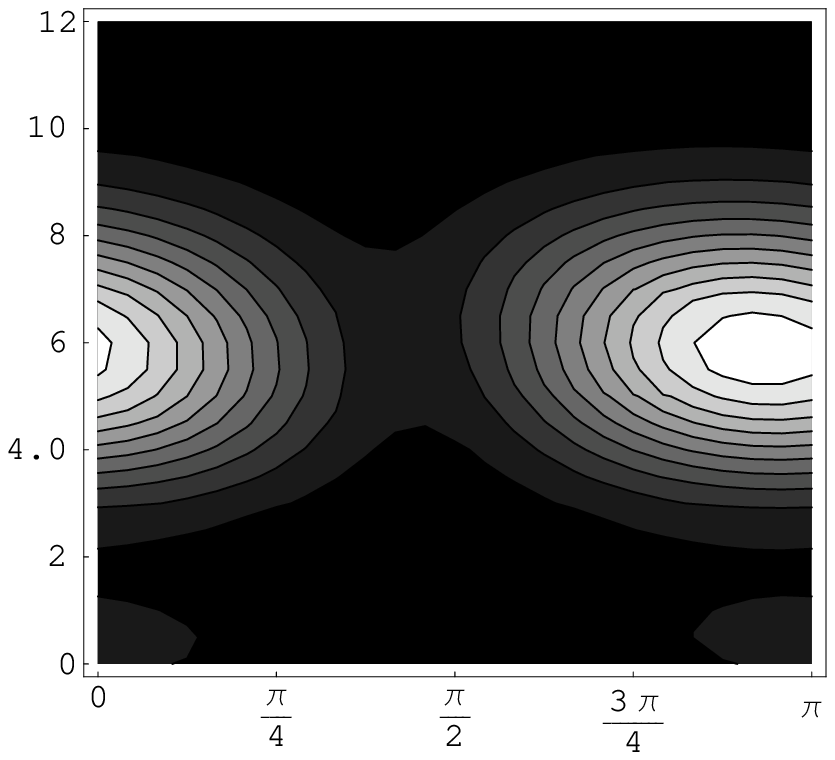}
\put(-100,-8){\large $\theta_1$} \put(-190,70){\large
$t$}\put(-20,150){(d)}
\end{center}
\caption{ The effect of different phases of the initial state of
the central qubit on the Fisher information in the resonance case
where $\omega_0=\omega_1=0.1$ and $\omega=0.1, g=0.5$ and
$\phi=\pi/4,~\pi/2$}
\end{figure}

From Fig.(1), one may conclude that,  at small values of the
rotating field strength, ( $\omega_1$) Fisher information has
periodic time  behavior, while at larger values of $\omega_1$, the
phenomena of the sudden changes are displayed. The number of peaks
and consequently the numbers of estimation regions increase as the
rotating field strength increases.
\begin{figure}[t!]
\begin{center}
\includegraphics[width=16pc,height=12pc]{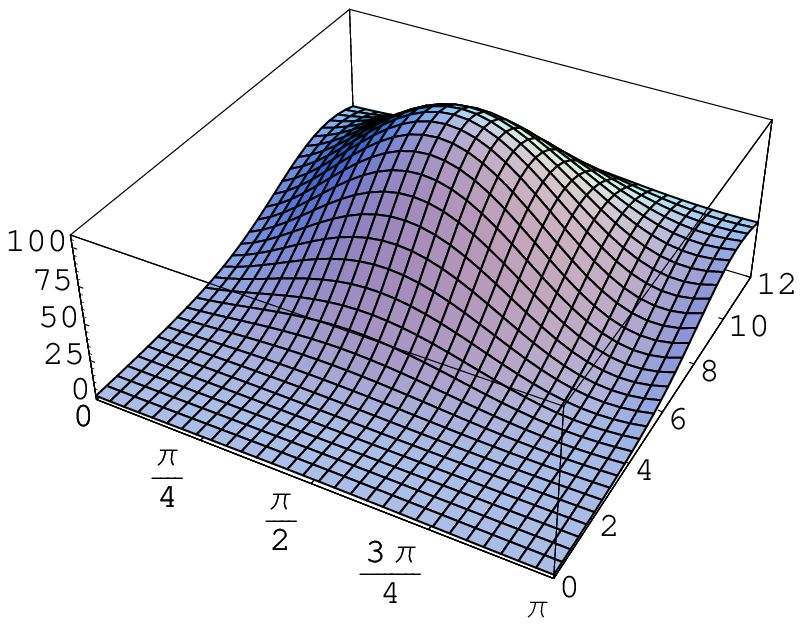}
\put(-215,70){\large $\mathcal{F}_{\theta_1}$}
\put(-145,12){\large$\theta_1$} \put(-20,40){\large
$t$}\put(-180,145){(a)}~~\quad
\includegraphics[width=16pc,height=12pc]{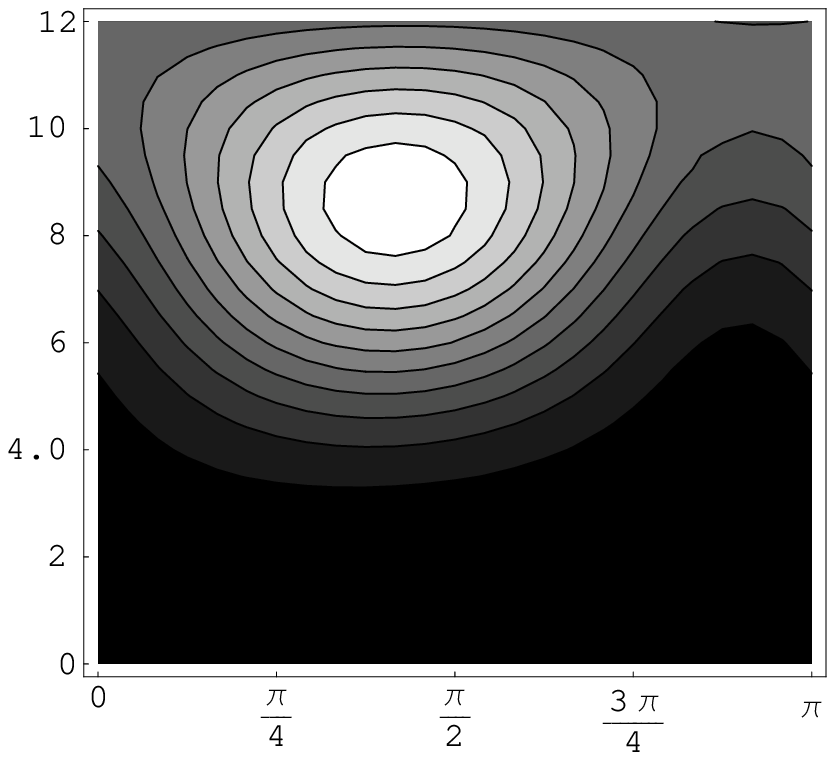}
\put(-100,-8){\large $\theta_1$} \put(-190,70){\large $t$}\put(-20,150){(b)}\\
\includegraphics[width=16pc,height=12pc]{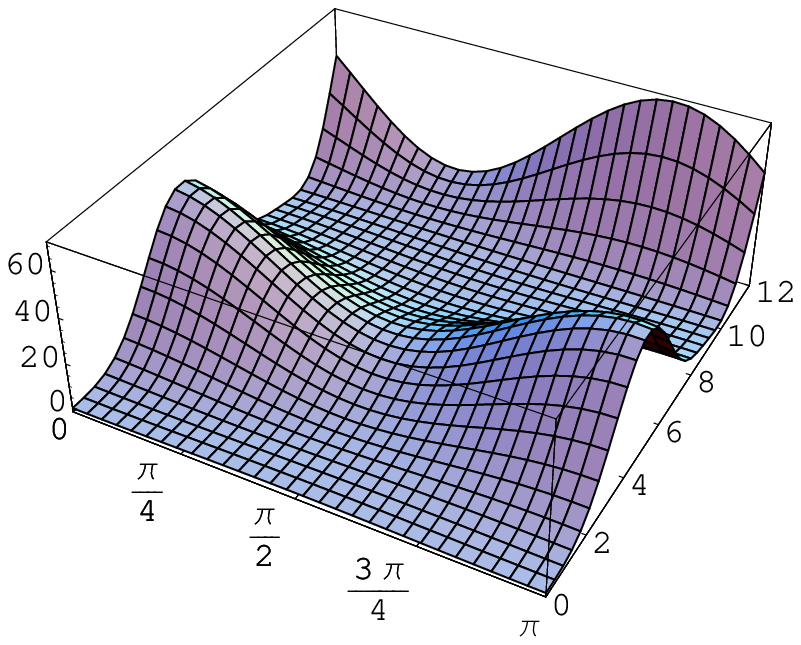}
\put(-215,70){\large $\mathcal{F}_{\theta_1}$}
\put(-145,12){\large$\theta_1$} \put(-20,40){\large
$t$}\put(-180,150){(c)}~~\quad
\includegraphics[width=16pc,height=12pc]{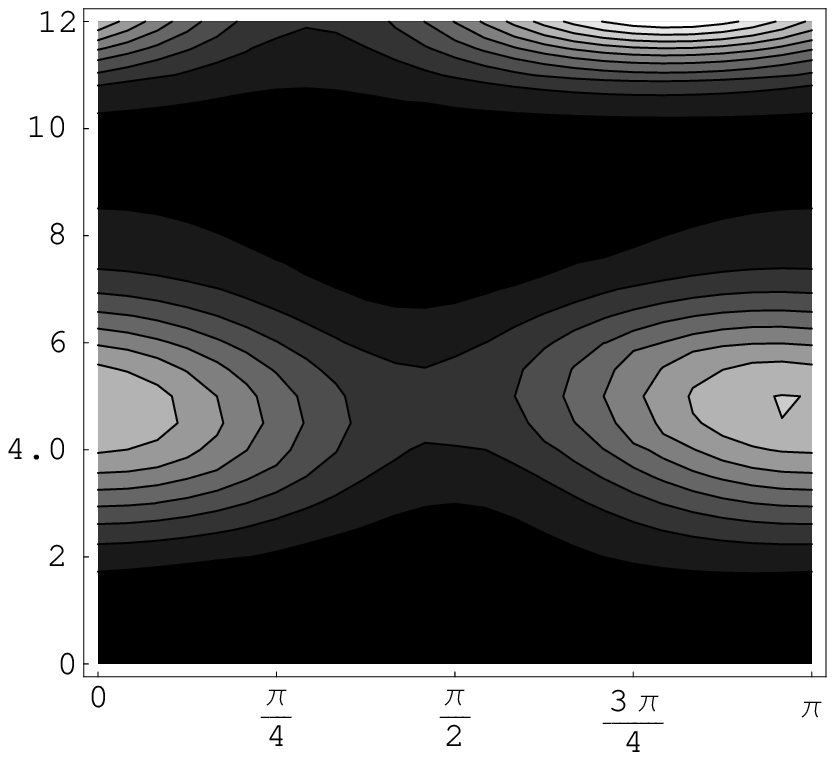}
\put(-100,-8){\large $\theta_1$} \put(-190,70){\large
$t$}\put(-20,150){(d)}
\end{center}
\caption{ The same as Fig1.(a,b) but for different values of the
coupling constant, where we set $g=0.3,0.7$ for (a,b) and (c,d),
respectively }
\end{figure}

The behavior of  Fisher information, ($\mathcal{F}_{\theta_1})$ in
the non-resonance  case is shown  in Fig.(2), where we fix  the
strength of the rotating field and the coupling constant. Two
different cases are considered; in the first  case, it is assumed
that, the longitudinal component of the field $(\omega_0)$ is
smaller than the transverse component $(\omega)$, while for the
second case we assume that, $(\omega_0>\omega)$. Our results
display that,  for the first case ($\omega_0<\omega$) the upper
bounds of the Fisher information are smaller than those depicted
for the second case ($\omega_0>\omega$). The minimum values of the
Fisher information $\mathcal{F}_{\theta_1}$ are observed in the
interval $\theta_1\in[\pi/2,3\pi/4]$, where the longitudinal
component is larger than the transverse component of the field,
while it is maximum for the second case, namely
($\omega_0>\omega)$.  The interaction time plays an important role
in both cases, where $\mathcal{F}_{\theta_1}$ increases as soon as
the interaction is switched on, while it takes longer time to
increase  for the second case. i.e., $(\omega>\omega_0)$.

Fig.(3) displays the behavior of the Fisher information
$\mathcal{F}_{\theta_1}$  for  different phases, where two values
of $\phi_1$ are considered and  have  used the same  values of the
parameters as in Fig.1(a,b). It is shown that, the upper bounds of
$\mathcal{F}_{\theta_1}$ increase as the phase angle increases.
Moreover, these upper bounds  of the Fisher information are
reached at small values of the  interaction time with  a large
phase angle. The size of estimation areas of the weight parameter
increases at the expense of the precision degree of estimation.

From Fig.(3), one may conclude that, the possibility of estimating
the weight parameter  $(\theta_1)$ increases  if  the initial
central qubit encodes only classical information, while this
possibility decreases  if  it encodes quantum information.

Fig.(4) displays the effect of different values of the coupling
constant,$(g)$ on  the behavior of the Fisher information
$\mathcal{F}_{\theta_1}$, and consequently on the estimation
degree of the weight parameter, $\theta_1$. It is clear that, at
small values of the coupling constant $(g)$, the maximum values of
$\mathcal{F}_{\theta_1}$ are larger than those displayed at larger
values of $g$.  The interaction time plays an important role on
the  Fisher information's behavior, where, at small values of the
coupling constant, $\mathcal{F}_{\theta_1}\simeq 0$ for
$t\in[0,2]$, then it increases for further values of interaction
time $t$. For larger values of the coupling constant, the maximum
peaks of the Fisher information are shifted, and consequently the
estimation areas are shifted at $\theta_1=0$ and $\pi$. The size
of the estimation areas increase at smaller values of $g$, while
the number of maximum peaks increases at larger values of the
coupling constant.

From Fig.(4), it is clear that, the estimation degree of the
weight parameter may be maximized  at smaller values of the
coupling constant and  the initial weight parameter, $\theta_1$
of the central qubit  is chosen in the interval
$\theta_1\in[\pi/4,~/3\pi/4]$ i.e, the central qubit codes quantum
 information. The other strategy, the estimation degree may be maximized  at larger values of the coupling constant and
 the initial  weight $\theta_1\in\Bigl\{[0,\pi/4]~ \mbox{or}~
 [3\pi/4,\pi]\Bigr\}$.

\begin{figure}[t!]
\begin{center}
\includegraphics[width=16pc,height=12pc]{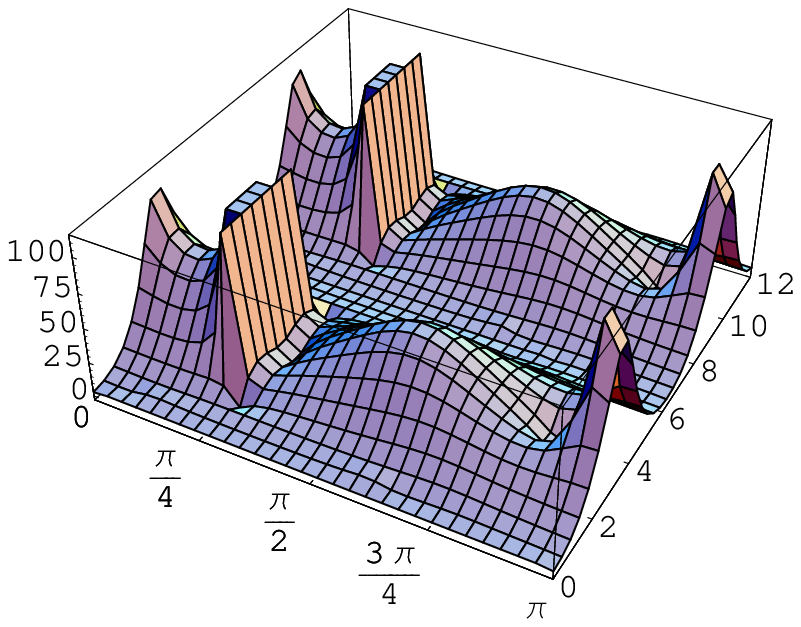}
\put(-205,70){\large $\mathcal{F}_{\theta_1}$}
\put(-145,12){\large$\theta_1$} \put(-20,40){\large
$t$}\put(-180,150){(a)}~~\quad
\includegraphics[width=16pc,height=12pc]{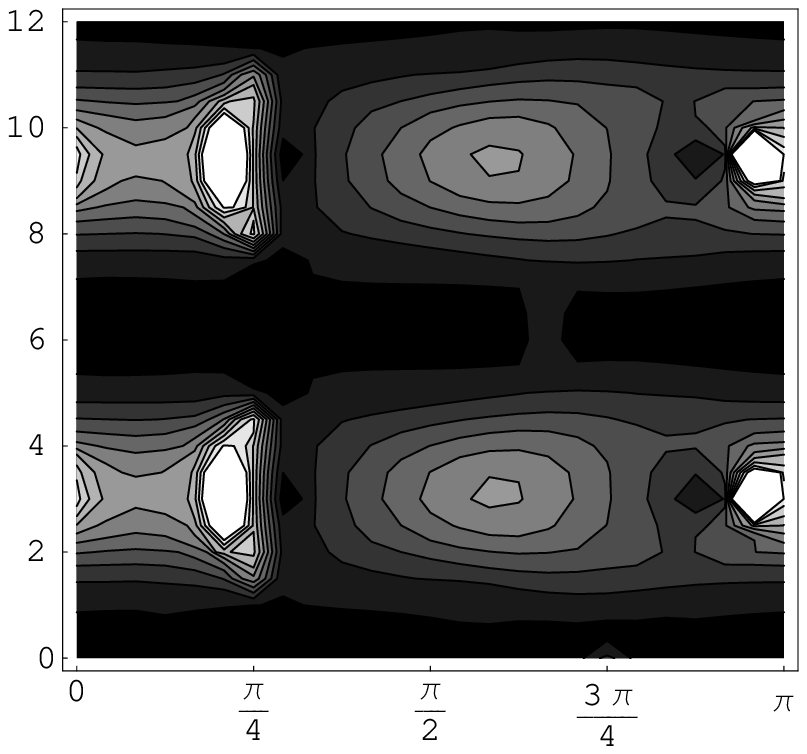}
\put(-90,-10){\large $\theta_1$} \put(-195,70){\large
$t$}\put(-20,150){(b)}\\
\includegraphics[width=16pc,height=12pc]{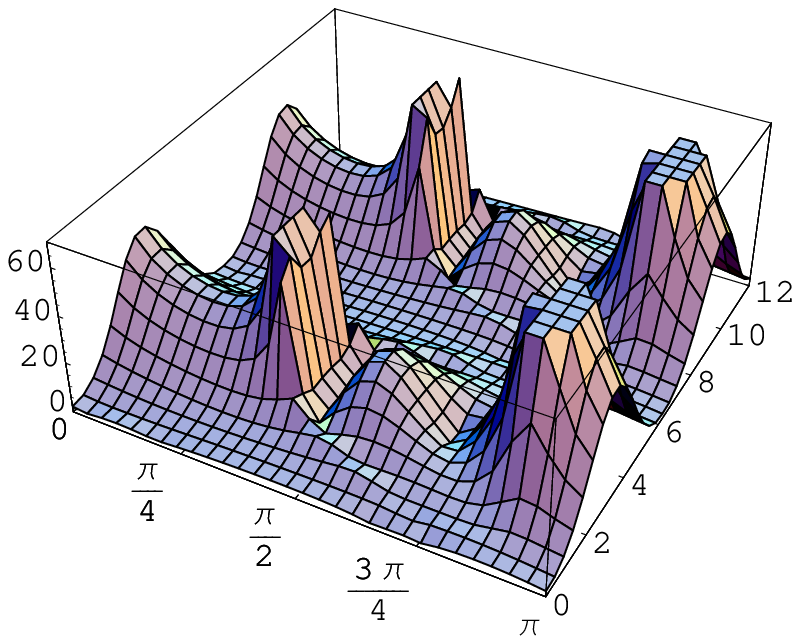}
\put(-205,70){\large $\mathcal{F}_{\theta_1}$}
\put(-145,12){\large$\theta_1$}
\put(-180,150){(c)}\put(-20,40){\large $t$}~~\quad
\includegraphics[width=16pc,height=12pc]{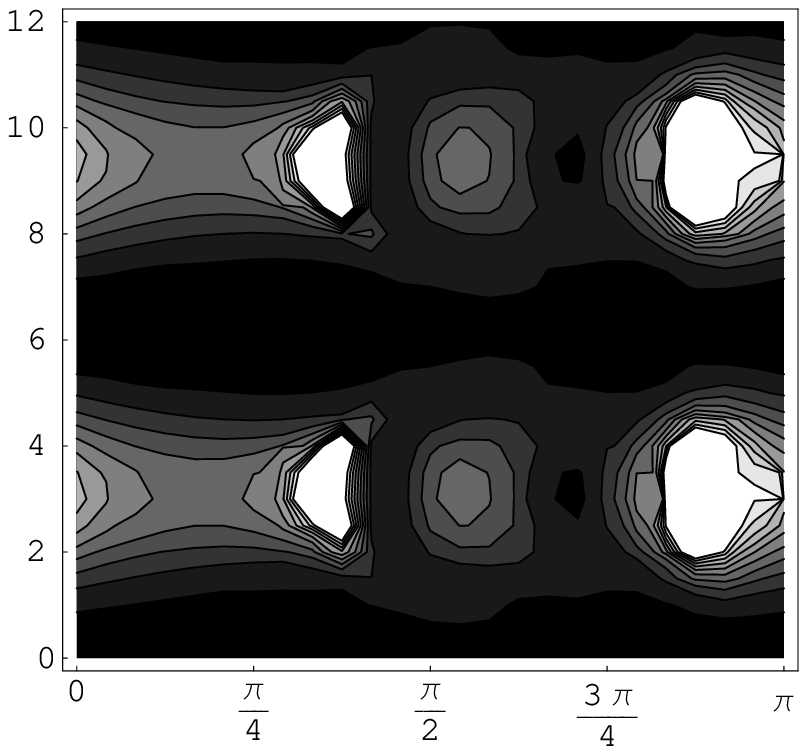}
\put(-90,-10){\large $\theta_1$} \put(-200,70){
$t$}\put(-20,150){(d)}\\
\includegraphics[width=16pc,height=12pc]{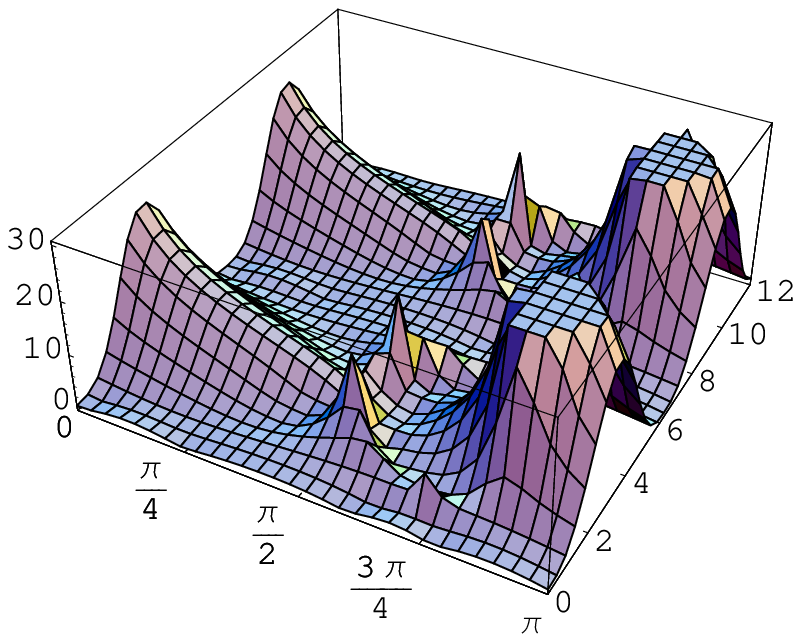}
\put(-205,70){\large $\mathcal{F}_{\theta_1}$}
\put(-145,12){\large$\theta_1$}
\put(-180,150){(e)}\put(-20,40){\large $t$}~~\quad
\includegraphics[width=16pc,height=12pc]{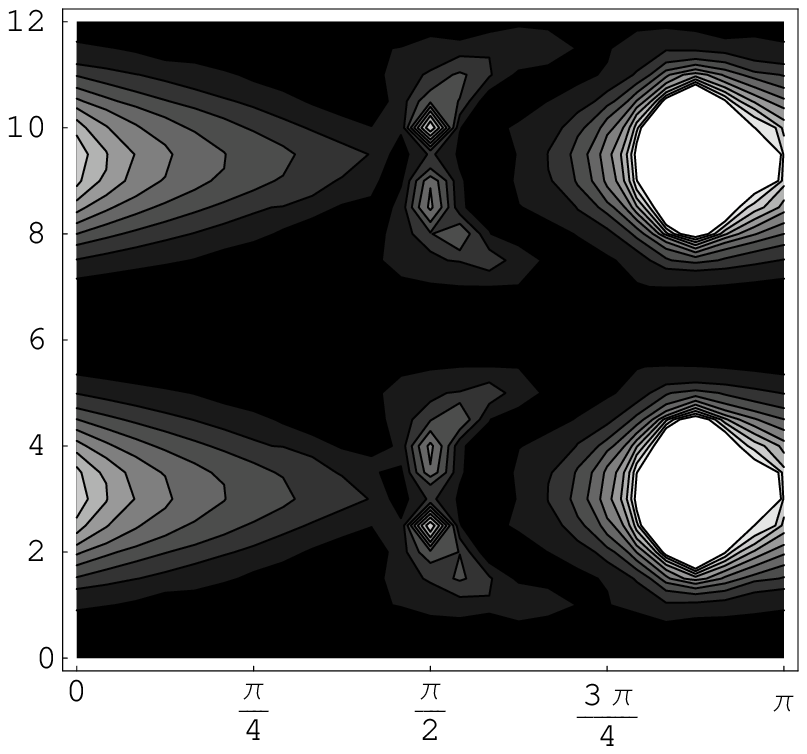}
\put(-90,-10){\large $\theta_1$} \put(-200,70){
$t$}\put(-20,150){(f)}
\end{center}
\caption{ Fisher information $\mathcal{F}_{{\theta_1}}$ against
for different initial  state settings of the spin qubit where
(a,b)~$\theta_2=\pi/2$, (c,d)~$\theta_2=\pi/3$ and
(e,f)~$\theta_2=\pi/4$,(g,h)~ $\theta_2=\pi/6$, while
$\phi_2=\pi,g=0.1$ and $\omega_1=0.5$.}
\end{figure}

 \begin{figure}
\begin{center}
\includegraphics[width=16pc,height=12pc]{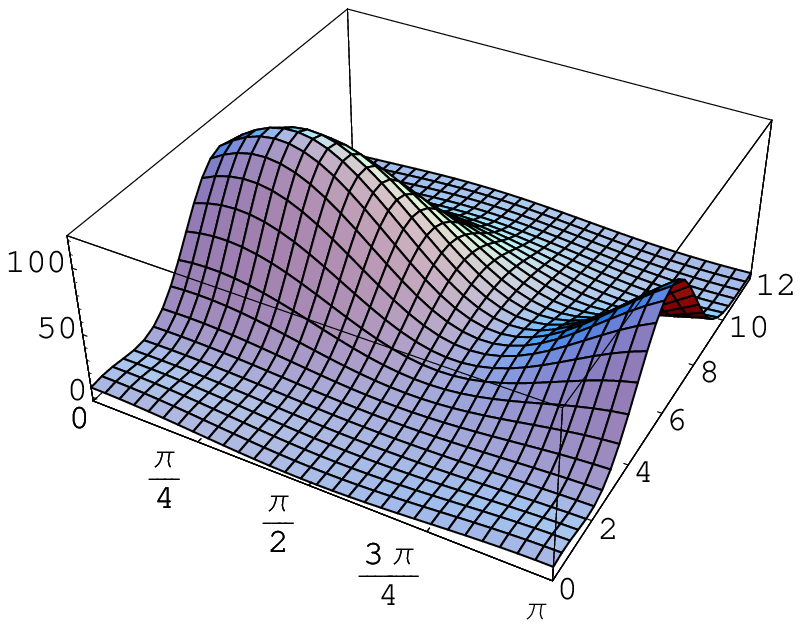}
\put(-205,70){\large $\mathcal{F}_{\phi_1}$}
\put(-145,12){\large$\theta_1$} \put(-20,40){\large
$t$}\put(-180,150){(a)}~~\quad
\includegraphics[width=16pc,height=12pc]{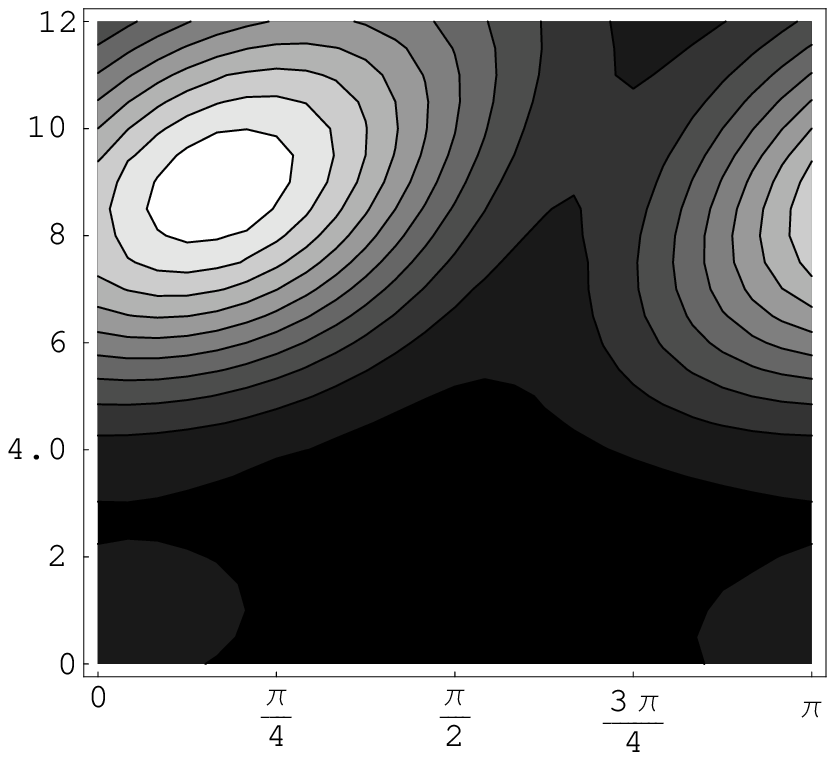}
\put(-100,-8){\large $\theta_1$} \put(-190,70){\large $t$}\put(-20,150){(b)}\\
\includegraphics[width=16pc,height=12pc]{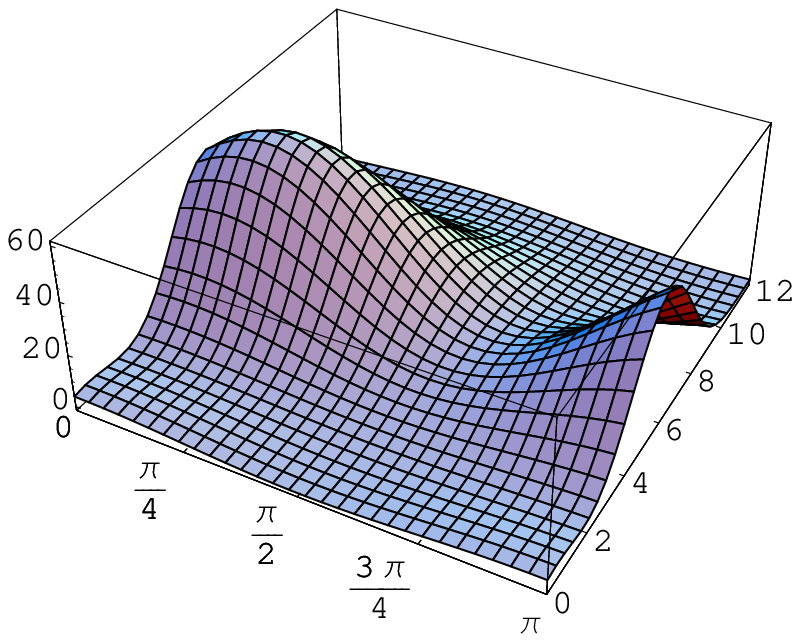}
\put(-205,70){\large $\mathcal{F}_{\phi_1}$}
\put(-145,12){\large$\theta_1$}
\put(-180,150){(c)}\put(-20,40){\large $t$}~~\quad
\includegraphics[width=16pc,height=12pc]{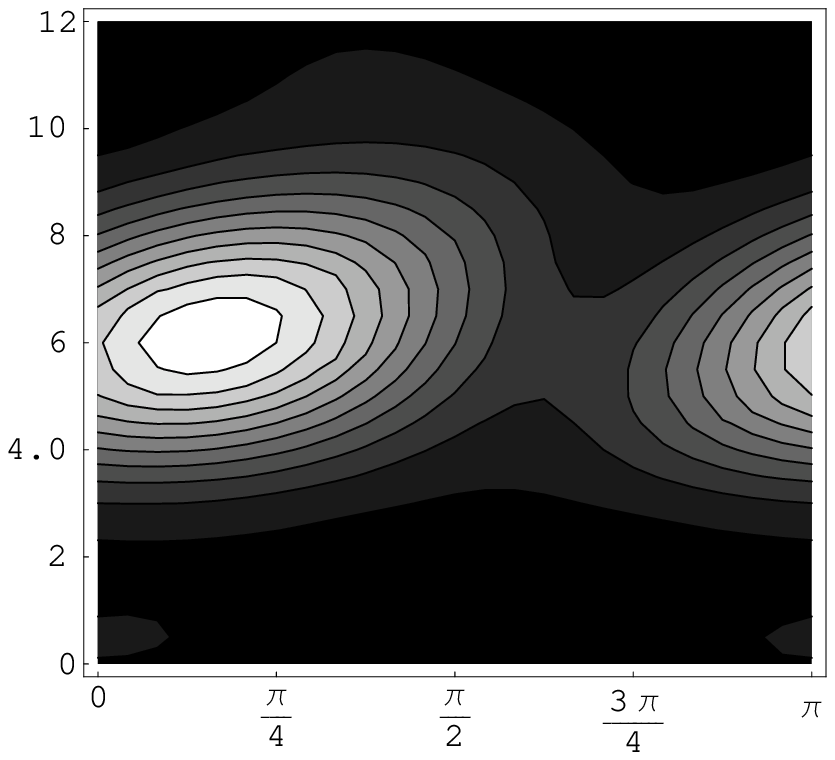}
\put(-100,-8){\large $\theta_1$} \put(-190,70){\large
$t$}\put(-20,150){(d)}
\end{center}
\caption{ Fisher information $\mathcal{F}_{\phi_1}$ against the
interaction time for resonant case and $\omega_1=g=0.5$ where (a)
at $\theta_1=\pi/2$ and (b) $\theta_1=\pi/4$.}
\end{figure}

 Fig.(5),  displays  the effect of  different initial state
settings of the spin-qubit on the quantum Fisher information with
respect to the weight  parameter $(\theta_1)$ of the central
qubit, where different values of the weight parameter $(\theta_2)$
are considered, while the phase parameter is fixed, namely,
$\phi_2=\pi$ . The general behavior shows that, the upper bounds
of $\mathcal{F}_{\theta_1}$ decreases as  $\theta_2$ decreases.
Also, as soon as the interaction is switched on, the Fisher
information increases gradually to reach its maximum value. For
further time  $\mathcal{F}_{\theta_1}$ decreases to reach its
minimum values. This behavior is repeated as the interaction time
increases to appear another peak. The areas in which one can
estimate the weight parameter with high precision  increase as
$\theta_2$  decreases.

From Fig.5, it is clear that if the spin-qubit encodes classical
information i.e.,($\ket{\psi_b}=-i\ket{1}$), the possibility of
estimating the weight parameter is larger than that depicted if it
encodes quantum information, where we set
$\ket{\psi_q}=\frac{1}{\sqrt{2}}(\ket{0}-i\ket{1}),~
\ket{\psi_q}=\frac{\sqrt{3}}{2}\ket{0}-\frac{i}{2}\ket{1}$ in
Figs.(5c,cd) and Figs.(5e,5f), respectively. Moreover, if the
spin-qubit encodes quantum information, the upper bounds depend on
the weight of the superposition of $\ket{0}$ and $\ket{1}$.

\section{Estimation the phase parameter}

 Fig.(6)  shows the behavior of the
Fisher information ($\mathcal{F}_{\phi_1}$) in the resonance case
with respect to the phase parameter $\phi_1$. Two different
initial state settings are considered, where we set the weight
parameter $\theta_1=\pi/2, \pi/4$, namely, the initial state of
the central qubit  is prepared in the state,
$\ket{\psi_q}=e^{-i\phi_1}\ket{1}$ and
$\ket{\psi_q}=\frac{1}{\sqrt{2}}(\ket{0}+e^{-i\phi_1}\ket{1})$,
respectively.
\begin{figure}
\begin{center}
\includegraphics[width=16pc,height=12pc]{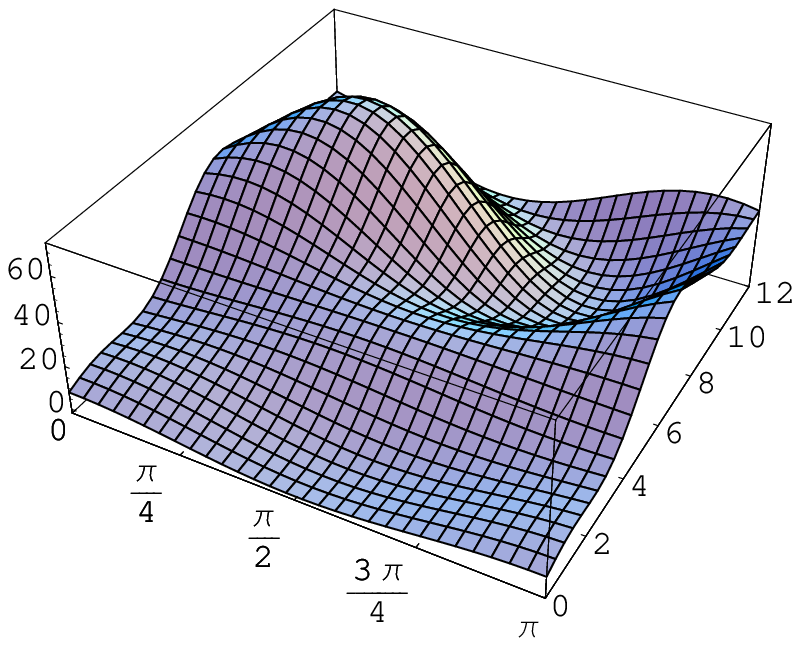}
\put(-215,70){\large $\mathcal{F}_{\phi_1}$}
\put(-145,12){\large$\theta_1$} \put(-20,40){\large
$t$}\put(-180,130){(a)}~~\quad
\includegraphics[width=16pc,height=12pc]{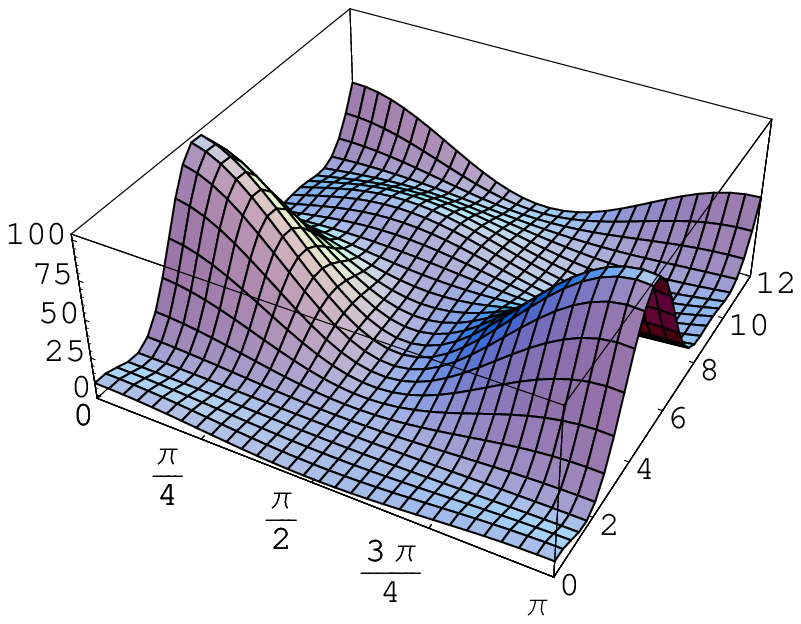}
\put(-205,70){\large $\mathcal{F}_{\phi_1}$}
\put(-145,12){\large$\theta_1$} \put(-20,40){\large
$t$}\put(-20,130){(b)}~~\quad
\end{center}
\caption{ Fisher information $\mathcal{F}_{\phi_1}$ against the
interaction time for the non-resonant case  with $\omega_1=g=0.5$
where (a) $\omega_0=0.09 $ and $\omega_1=0.01$  and (b)
 $\omega_0=0.01 $ and $\omega_1=0.09$.}
\end{figure}
The results  show that, the behavior of the Fisher information  is
similar for the two different initial states, but only the upper
bounds at $\theta_1=\pi/2$ are larger than those displayed at
$\theta_1=\pi/4$.  In the non-resonance case, the behavior depends
on which component is larger. However, if  the longitudinal
component, $\omega_0$ is larger than the transverse component
$\omega$, then  the upper bounds are smaller than those displayed
for $\omega>\omega_0$. Moreover, the upper and lower bounds of
$\mathcal{F}_{\phi_1}$ appear at different values of interaction
time. These results are displayed in Fig.(7) where it is assumed
that, the initial state of the central qubit is prepared by
setting $\theta_1=\pi/2$.

The effect of the coupling constant $g$ is displayed in Fig.(8),
where we consider the non-resonance case  and a fixed value of the
rotating field component. It is clear that, at smaller values of
the coupling constant the upper bounds of  the Fisher information
are smaller than those displayed at large coupling constant. On
the other hand, the estimated areas are larger for smaller values
of the coupling constant. The estimating area for  larger values
of the coupling constant appears at smaller values of the
interaction time, while it takes longer time  to appear  at small
values of the  coupling constant, $g$.

\begin{figure}[t!]
\begin{center}
\includegraphics[width=16pc,height=12pc]{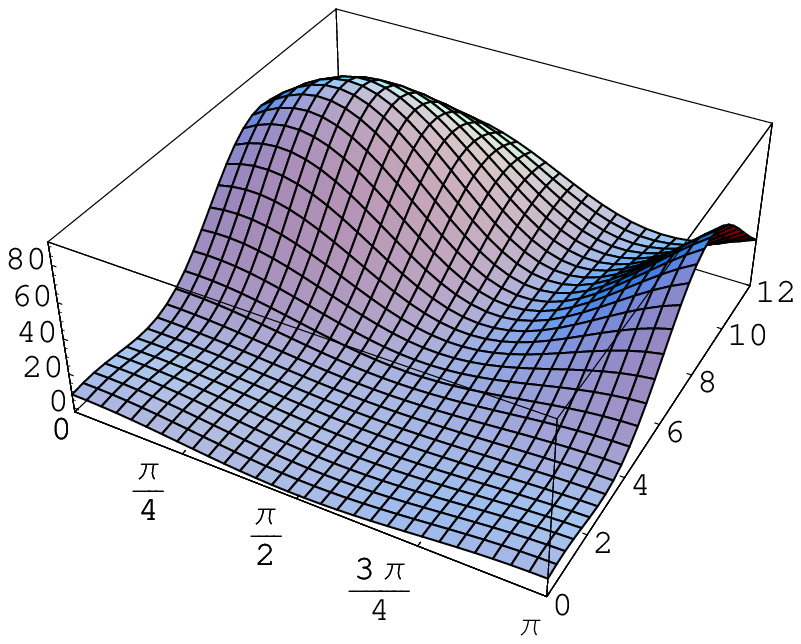}
\put(-145,12){\large$\theta_1$} \put(-20,40){\large
$t$}\put(-180,130){(a)}~~\quad\quad
\includegraphics[width=16pc,height=12pc]{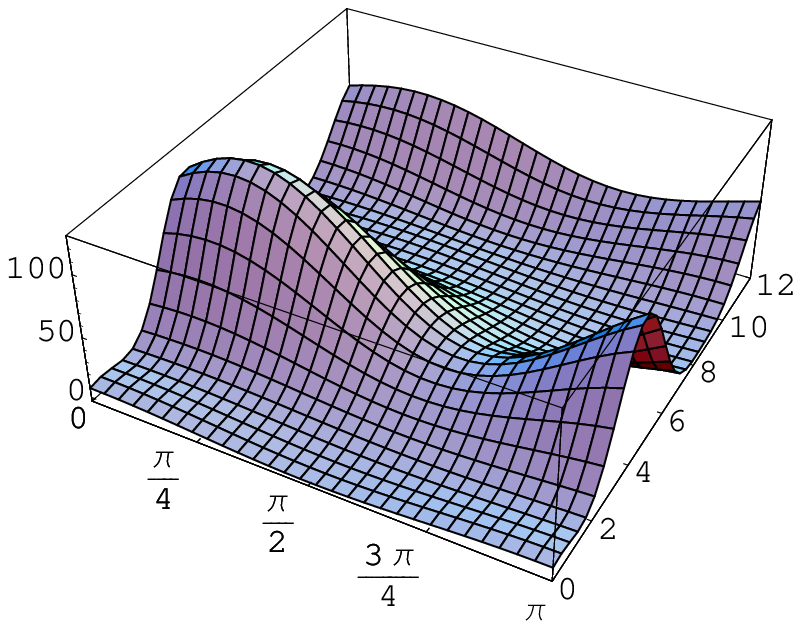}
\put(-210,70){\large $\mathcal{F}_{\phi_1}$}
\put(-145,12){\large$\theta_1$} \put(-20,40){\large
$t$}\put(-20,130){(b)}~~\quad
\end{center}
\caption{ Fisher information $\mathcal{F}_{\phi_1}$ against the
interaction time at $\Delta=0$ and $\omega_1=g=0.5$ where (a)
$g=0.3$, (b) $g=0.7$.}
\end{figure}

From Figs.(6-8), one may conclude that, for the resonance case the
precision of estimating the phase parameter depends on the initial
state settings, where if the central qubit is prepared in a state
encodes classical information, the possibility of the estimation
degree of the phase parameter is larger than that depicted if the
central qubit encodes quantum information. For the non-resonance
case, the estimation degree depends on which component
(longitudinal/ transverse) of the field is larger. Larger values
of the coupling constant increase the  size of the  estimation
area, in which one may estimate the phase parameter with high
degree of estimation.

\begin{figure}[t!]
\begin{center}
\includegraphics[width=16pc,height=12pc]{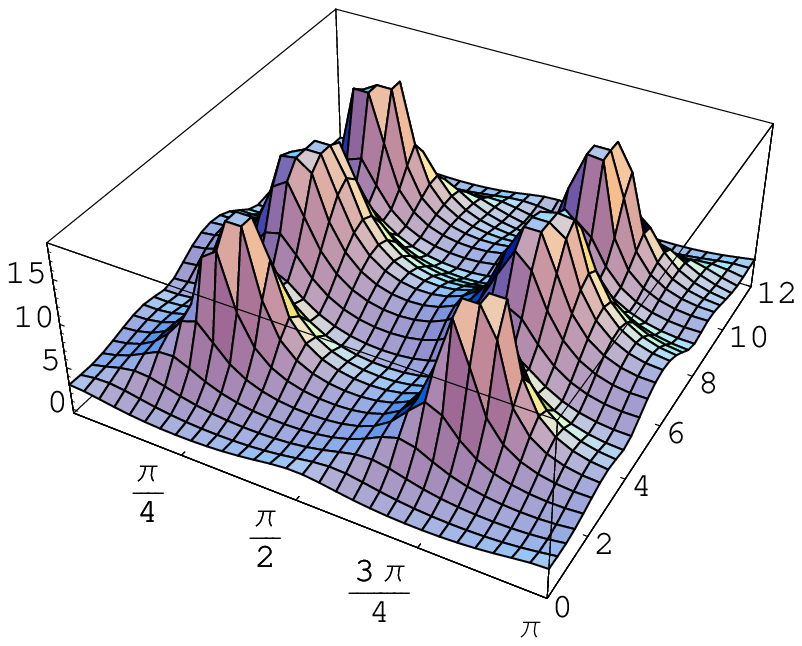}
\put(-215,70){\large $\mathcal{F}_{\phi_1}$}
\put(-145,12){\large$\theta_1$} \put(-20,40){\large
$t$}\put(-180,130){(a)}~~\quad\quad
\includegraphics[width=16pc,height=12pc]{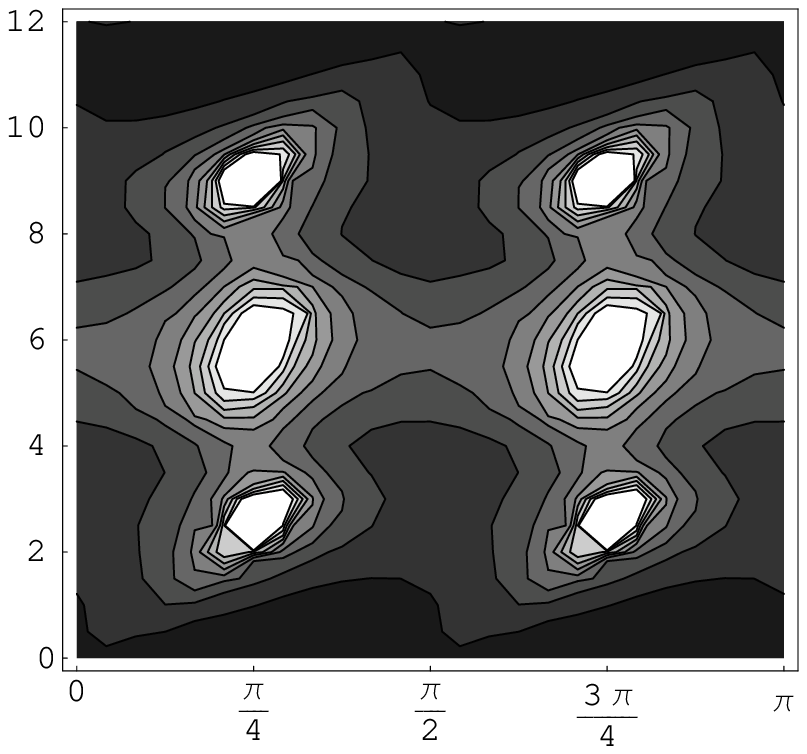}
\put(-195,70){\large $t$} \put(-95,-10){\large$\theta_1$}
\put(-20,40){\large $t$}\put(-20,147){(b)}~~\quad\\
\vspace{0.2cm}
\includegraphics[width=16pc,height=12pc]{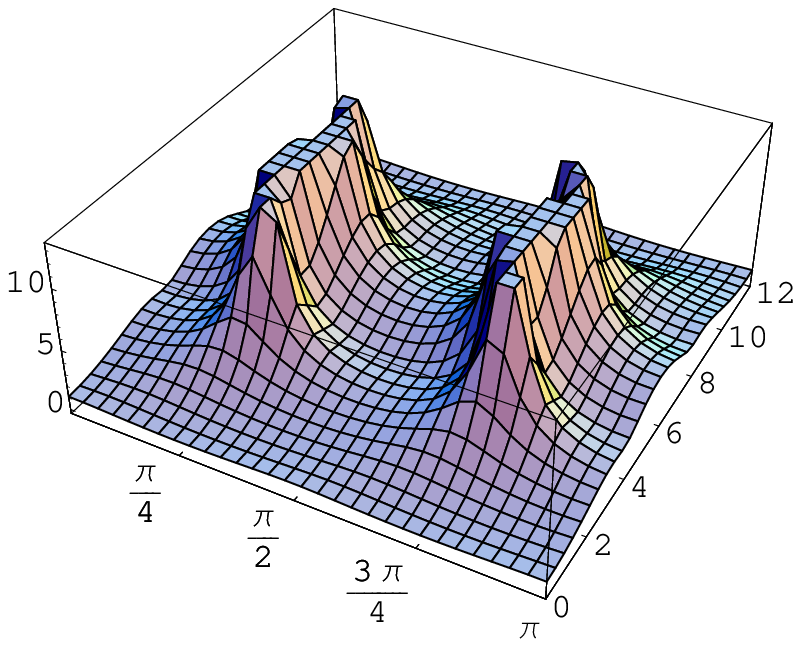}
\put(-215,70){\large $\mathcal{F}_{\phi_1}$}
\put(-145,12){\large$\theta_1$} \put(-20,40){\large
$t$}\put(-180,130){(c)}~~\quad\quad
\includegraphics[width=16pc,height=12pc]{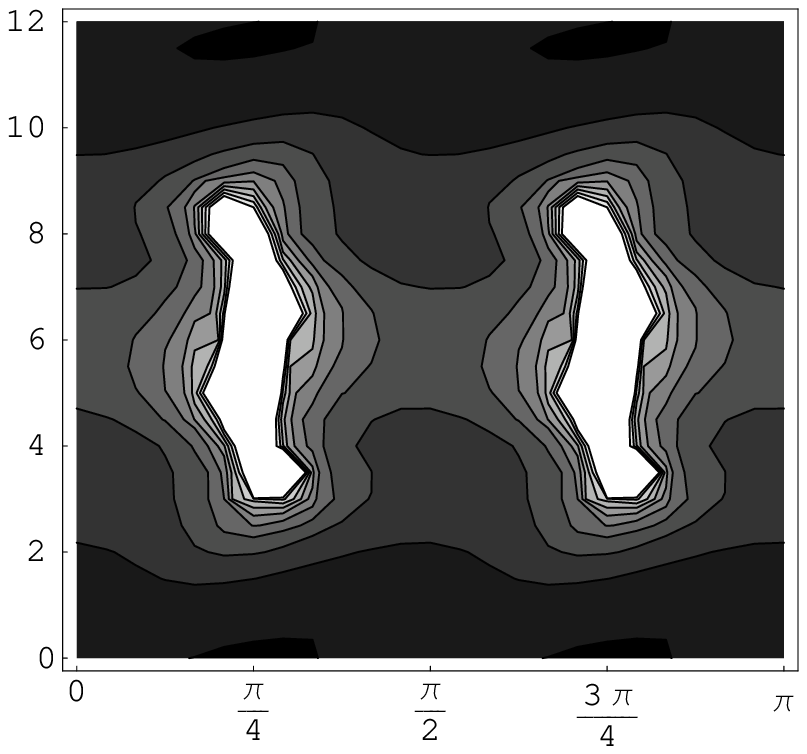}
\put(-195,70){\large $t$} \put(-95,-10){\large$\theta_1$}
\put(-20,40){\large
$t$}\put(-20,147){(d)}~~\quad \\
\vspace{0.2cm}
\includegraphics[width=16pc,height=12pc]{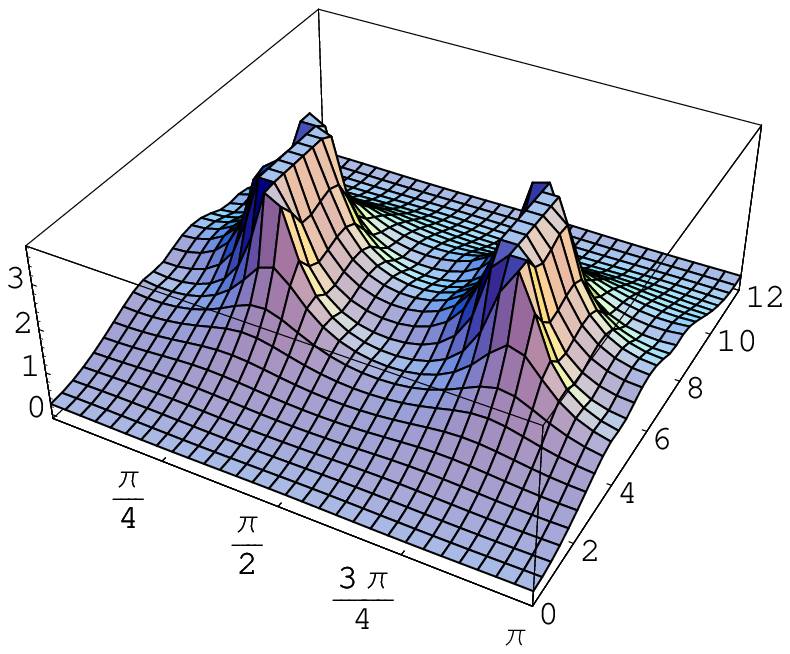}
\put(-215,70){\large $\mathcal{F}_{\phi_1}$}
\put(-145,12){\large$\theta_1$} \put(-20,40){\large
$t$}\put(-180,130){(e)}~~\quad\quad
\includegraphics[width=16pc,height=12pc]{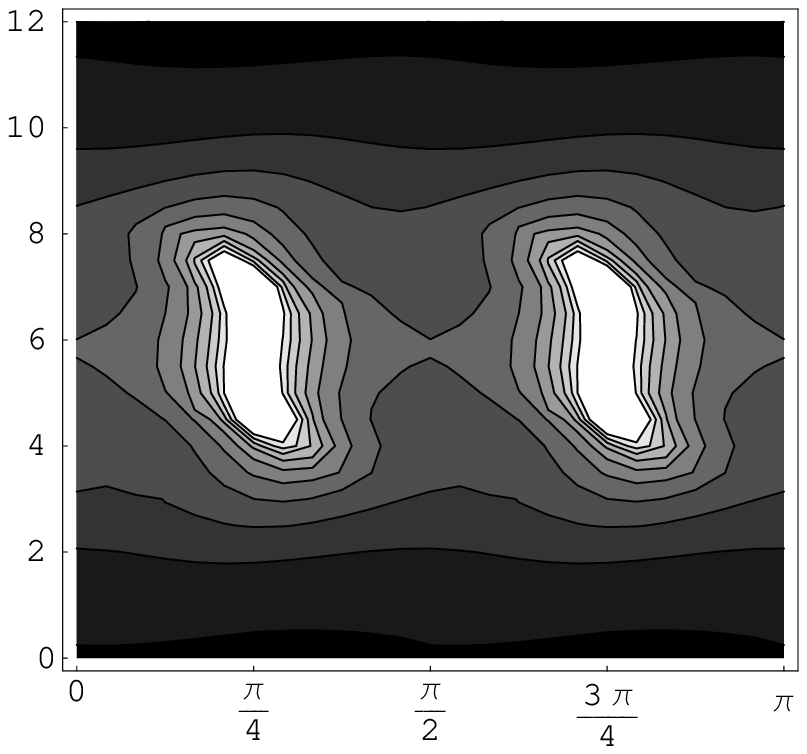}
\put(-195,70){\large $t$} \put(-95,-10){\large$\theta_1$}
\put(-20,40){\large $t$}\put(-20,147){(f)}~~\quad

\end{center}
\caption{ Fisher information $\mathcal{F}_{\phi_1}$ against the
interaction time at $\Delta=0$ and $\omega_1=0.1,
g=0.5,~\theta_1=\frac{\pi}{2},\phi_2=\pi$, where (a,b)
$\theta_2=\frac{\pi}{2}$ (c,d) $\theta_2=\frac{\pi}{4}$ (e,f)
$\theta_2=\frac{\pi}{6}$.}
\end{figure}

Figs.(9), display the behavior of $\mathcal{F}_{\phi_1}$ for
different initial states of the spin-qubit. It is clear that, for
larger values of $\theta_2$  the upper bounds are larger than
those displayed  at smaller values of $\theta_2$.  It is clear
that, $\mathcal{F}_{\phi_1}$ increases as $\theta_1$ increases to
reach its maximum values at $\theta_2=\frac{\pi}{4}$. On the other
hand,  the minimum values of $\mathcal{F}_{\phi_1}$ are depicted
at $\theta_1=\frac{\pi}{2}$. Moreover, the number of peaks  and
the estimated areas increases as the weight parameter
 of the spin-qubit increaser.

 From Figs.(5) and (9), it is clear that the initial state
 settings of the spin-qubit has a clear effect on the
 precision degree of estimating the weight and the phase
 parameters of the central qubit.

\section{Spin Bath problem}
In this context, it is important to mention that in the previous
sections,  we consider one
 central qubit interacting with  a minimum dimension of the bath,
namely we consider {\it a single} spin-qubit, which is polarized
in $z-$ direction. In this case, the state of the spin particle is
defined by $\rho_s=\frac{1}{2}(1\pm \tau_z)$. If we have a bath of
large number of particles  polarized in $z$-direction, then their
initial state is defined by $\rho_s(N)=\frac{1}{2^N}(I_{2\times
2}\pm \tau_z)^{\otimes N}$, which is more complicated to solve.
Therefore, to simplify the calculations, we assume that the $N$
spin particles are unpolarized. Under this condition, the states
of the $N$-bath unpolarized particles reduces to be
$\rho_s(N)=\frac{1}{2^N}I_{2\times 2}$. However,  a similar
calculation may be done, where the detuning parameter is affected
to be $-\frac{1}{2}gN\leq\Delta\leq\frac{1}{2}gN$ and consequently
$\Delta _{\pm}=\Delta \pm\frac{gN}{2}$, where it is assumed that
$g_1=g_2=.....=g_N=g$.

\begin{figure}[t!]
\begin{center}
\includegraphics[width=16pc,height=12pc]{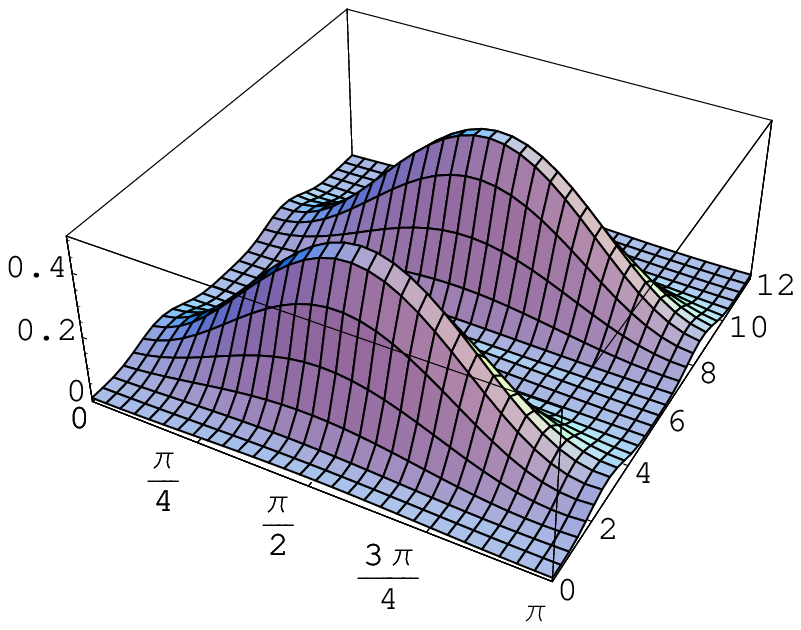}
\put(-215,70){\large $\mathcal{F}_{\theta_1}$}
\put(-145,12){\large$\theta_1$} \put(-20,40){\large
$t$}\put(-180,130){(a)}~~\quad\quad
\includegraphics[width=16pc,height=12pc]{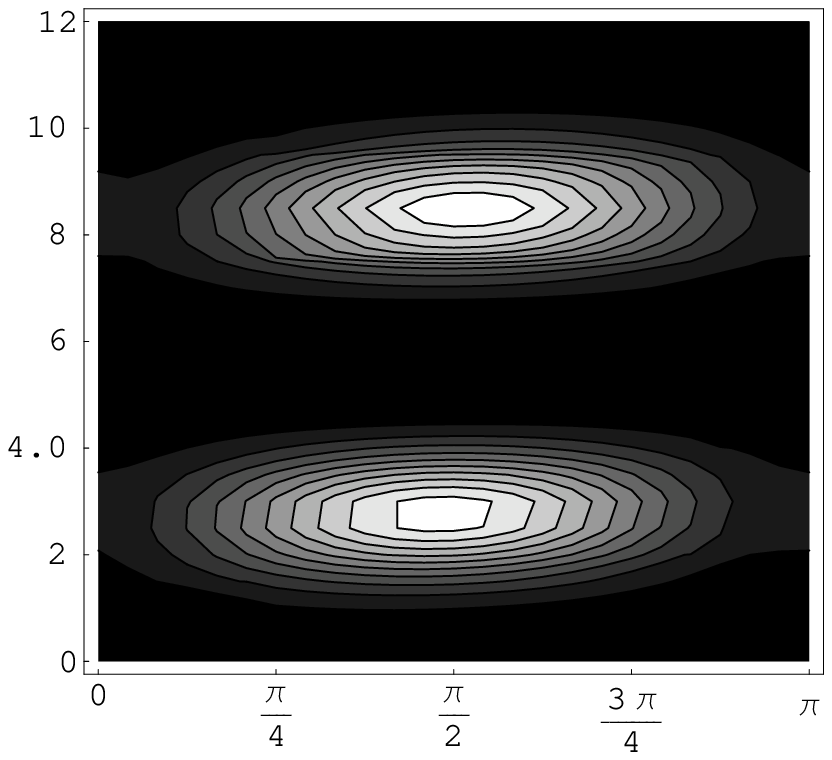}
\put(-190,70){\large $t$} \put(-90,-10){\large$\theta_1$}
\put(-20,40){\large
$t$}\put(-20,147){(b)}~~\quad\\
\vspace{0.2cm}
\includegraphics[width=16pc,height=12pc]{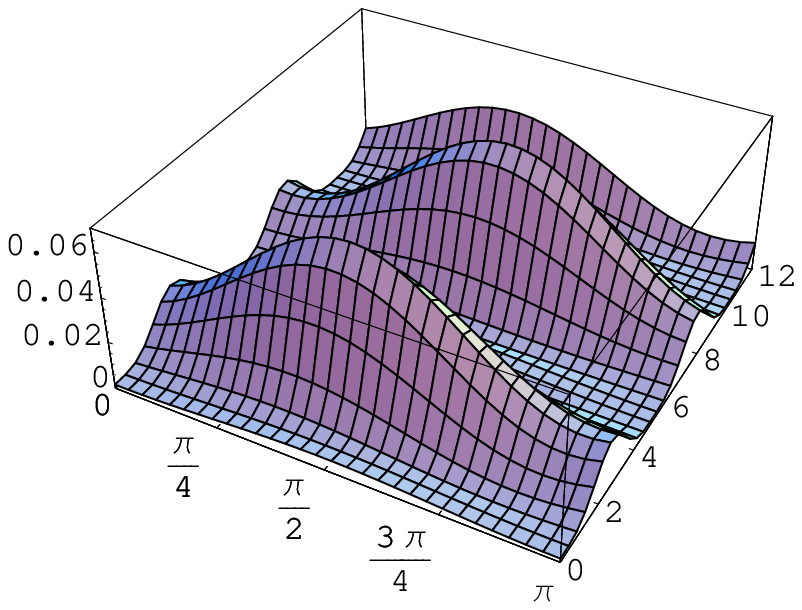}
\put(-215,70){\large $\mathcal{F}_{\theta_1}$}
\put(-145,12){\large$\theta_1$} \put(-20,40){\large
$t$}\put(-180,130){(c)}~~\quad\quad
\includegraphics[width=16pc,height=12pc]{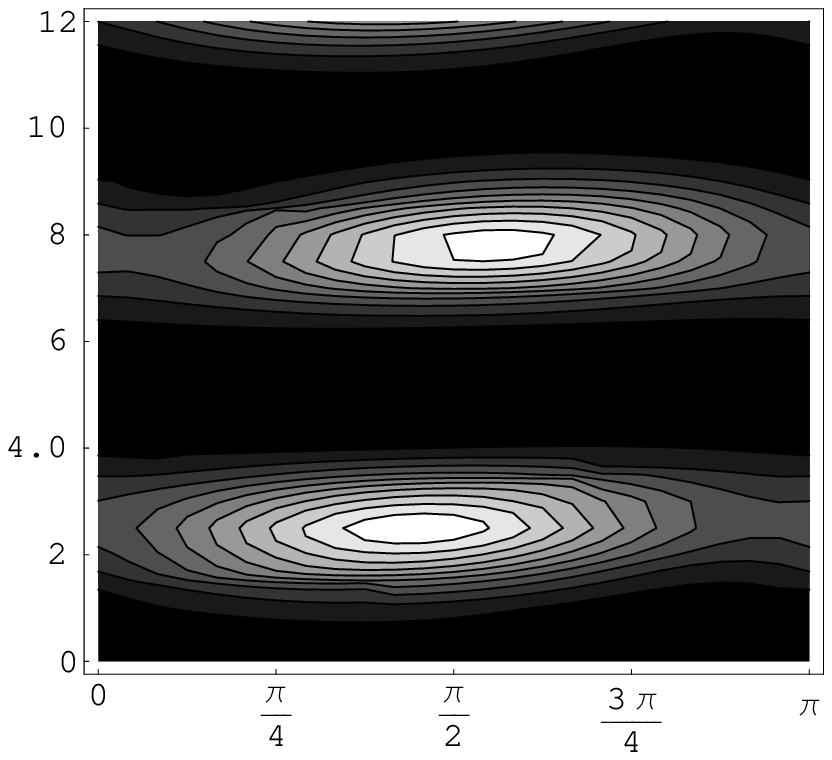}
\put(-190,70){\large $t$} \put(-90,-10){\large$\theta_1$}
\put(-20,40){\large
$t$}\put(-20,147){(d)}~~\quad\\

\end{center}
\caption{ Fisher information $\mathcal{F}_{\theta_1}$ against the
interaction time at $\Delta=0$ and $\omega_1=0.5,g=0.1$ where (a)
$N=5$, (b) $N=7$.}
\end{figure}
In Fig.(10), we consider  the central qubit interacts with larger
number of unpolarized spin-bath particles.  The general behavior
shows that, the upper bounds decrease as the numbers of spin-bath
particle increase, where  we consider that $N=5, 7$ particles in
Figs.(10a,10b) and (10c,10d,) respectively. As the initial weight
parameter increases, $\mathcal{F}_{\theta_1} $ increases to reach
its maximum value at $\theta_1=\pi/2$, then decays gradually to
reach its minimum   value at $\theta_1=\pi$. Moreover, the upper
bounds are shifted as the interaction time increases to appear at
a different value of the initial weight. These results are clearly
described on Figs.(10a-10d).

\section{ Conclusion}
In this contribution, Fisher information  is used as an estimator
of the  weight  and the phase parameters of a central qubit
interacts with a single- qubit in the presence of a magnetic
field.
 The effect of the magnetic field parameters
on the degree of estimating the weight and phase  parameters is
investigated. For resonance case, the  possibility of estimating
the weight parameter may be achieved with high degree of precision
at large values of the rotating field strength. Also, the size of
the areas, in which one may estimate the weight parameter,
decreases at the expense of their numbers, where the numbers of
these areas increase as the rotating field strength increases. For
non-resonance case, the upper bounds depend on the strength of the
field longitudinal/transverse parameters. It is clear that, the
upper bounds are larger if the transverse strength is  stronger
than the longitudinal strength. The possibility of estimating  the
weight parameter increases if it encodes only classical
information. The coupling constant plays an important role in
controlling the estimation precision, where estimation degree is
large at smaller values of the coupling constant. On the other
hand,  a larger interaction time  is required  in order to
estimate the weight parameter. Further, a shorter time of
interaction is required if the coupling constant is larger.

The behavior of Fisher information with respect to the phase
parameter is discussed for different values of the initial state
setting of the central qubit. It is shown that, for the resonance
case, the possibility of estimating the phase parameter increases
if the initial state  of the central qubit encodes quantum
information. For the non-resonance  case, the estimation degree of
the phase parameter depends on the components of the magnetic
field. For large values of the coupling constant, the upper bounds
of estimation degree of the phase parameter, increases at the
expense of the size of the estimation areas.

The effect of different initial state settings of the spin- qubit
on the estimation degree of the weight and the phase parameters is
discussed. It is shown that, Fisher information corresponding to
these two parameters decreases as the weight parameter of the
spin-qubit decreases.

The Fisher information with respect to the  weight parameter of
the  central qubit interacts with unpolarized  large number of
spin- bath particles is discussed. We show that, the upper pounds
of the Fisher information decreases as the bath-spin numbers
increase.As the interaction time increases, the upper bounds
appear at larger values of initial weight of the central qubit.

{\bf Acknowledgements\\} I would like to thank the referee for
his/her important comments which helped to improve the manuscript.
Also, I thank Prof.  S.S. Hassan for his valuable suggestions
which modified our discussion

\end{document}